\documentclass[twocolumn,showpacs]{revtex4}

\usepackage[dvips]{graphicx}
\usepackage{color}

\begin{document}

\title{Analysis for practical realization of number-state manipulation \\
by number-sum Bell measurement with linear optics}
\date{\today}
\author{Akira Kitagawa}
\altaffiliation{E-mail address: kitagawa@nict.go.jp \\
Address after April, 2004:
National Institute of Information and Communications Technology,
4-2-1 Nukui-Kita, Koganei, Tokyo 184-8795, Japan
}
\author{Katsuji Yamamoto}
\altaffiliation{E-mail address: yamamoto@nucleng.kyoto-u.ac.jp}
\affiliation{Department of Nuclear Engineering,
Kyoto University, Kyoto 606-8501, Japan
}

\begin{abstract}
We analyze the linear optical realization
of number-sum Bell measurement and number-state manipulation
by taking into account the realistic experimental situation,
specifically imperfectness of single-photon detector.
The present scheme for number-state manipulation
is based on the number-sum Bell measurement,
which is implemented with linear optical elements, i.e.,
beam splitters, phase shifters and zero-one-photon detectors.
Squeezed vacuum states and coherent states are used as optical sources.
The linear optical Bell state detector is formulated quantum theoretically
with a probability operator measure.
Then, the fidelity of manipulation and preparation
of number-states, particularly for qubits and qutrits,
is evaluated in terms of the quantum efficiency and dark count
of single-photon detector.
It is found that a high fidelity is achievable
with small enough squeezing parameters and coherent state amplitudes.
\end{abstract}

\pacs{03.67.Mn, 03.67.Hk, 42.50.Dv}

\maketitle

%%%%%%%%%%
\section{Introduction}
\label{sec:introduction}
%%%%%%%%%%

Extensive research and development
have been done recently on quantum information and communication
technologies.
Among various media for quantum information and communication,
the photon-number Fock space is promising
in the point that it provides higher dimensional states
such as qutrits to carry more information than qubits.
This stimulates great interest in preparation and manipulation
of various photon-number states.
Specifically, teleportation \cite{BBCJPW,Braunstein}
is known to provide important tools
for quantum communication and information processing.
The number-state teleportation may be performed
by making a number-sum Bell measurement
with certain Einstein-Podolsky-Rosen (EPR) resource
\cite{BBCJPW,Milburn}.
Then, its method really appears to be useful
for engineering the input states, irrespective of teleportation fidelity.
In fact, a quantum scissors for number-state truncation
by projective measurement, which has been investigated thoroughly so far
\cite{scissors1,scissors2,scissors3},
may be viewed as a teleportation-based number-state manipulation.
The entanglement resource is prepared
from vacuum and one photon state through a 50:50 beam splitter,
and the joint photon detection implements the number-sum Bell measurement.
An experimental realization of quantum scissors has been done recently,
generating a qubit of vacuum and one-photon state
by truncating a coherent state \cite{scissors4}.
It is also interesting that an experimental result
has been reported for the teleportation of the vacuum-one-photon qubit
\cite{LSPM}.

The number-sum Bell measurement accordingly plays an essential role
for engineering the photon-number states via teleportation.
Some feasible schemes have appeared recently
for implementing particularly
the joint measurement of number-sum and phase-difference with linear optics
\cite{pdm1,KY-2002,KY-2003,ZPM-2003,ZPM-2004},
and an experimental demonstration has also been reported
\cite{pdm2}.
Then, various number-state preparations and manipulations
have been investigated based on teleportation
with number-sum Bell measurements and relevant EPR resources
\cite{KY-2002,KY-2003,ZPM-2003,ZPM-2004}.
In these respects, there are growing interests
in the number-sum Bell measurement
and its application for the number-state manipulation.

In this paper we analyze the linear optical realization
of number-sum Bell measurement and number-state manipulation
by taking into account the realistic experimental situation,
specifically imperfectness of single-photon detector.
The present scheme for number-state manipulation
is based on the number-sum Bell measurement,
which is implemented with linear optical elements,
i.e., beam splitters, phase shifters and zero-one-photon detectors.
As for the optical sources,
many useful manipulations of number-states are realized
with squeezed vacuum states and coherent states,
which are widely used in optical experiments,
while single-photon sources may not be required
\cite{KY-2002,KY-2003,ZPM-2003,ZPM-2004}.
Beam splitters and phase shifters will be available with high accuracy.
On the other hand, photon detectors are currently developed devices,
which in practice have finite quantum efficiency
and nonzero dark count rate.
Hence, for feasible experiments it is desired
to provide a systematic method to evaluate the efficiency
of number-state manipulation with number-sum Bell measurement,
by taking into account the imperfectness of actual photon detectors.
It is indeed encouraging
that some significant developments and new proposals have been made
for single-photon detection to achieve the quantum efficiency
close to unity \cite{SPD1,SPD2}.
We believe that the present work promotes
future experimental efforts on engineering photon-number states
by number-sum Bell measurement.

This paper is organized as follows.
In Sec. \ref{sec:bell-state}, we describe the number-sum Bell states,
particularly those associated with phase-difference.
In Sec. \ref{sec:bell-detector},
we present a linear optical detector
to measure a specific number-sum Bell state,
and formulate it quantum theoretically
with a probability operator measure (POM).
Then, we estimate the sensitivity of these detectors
in terms of the efficiency of practical single-photon detectors.
In Sec. \ref{sec:manipulation}, we investigate the number-state manipulation
via teleportation by number-sum Bell measurement.
We present the formulas to evaluate the fidelity
for engineering various photon-number states.
In Sec. \ref{sec:analysis}, by applying these formulas
we analyze the efficiencies of some useful manipulations and preparations
in particular for qubits and qutrits.
This analysis indicates that these experiments will be performed
with good fidelities by utilizing currently available apparatus.
Section \ref{sec:summary} is devoted to summary.

%%%%%%%%%%
\section{Number-sum Bell states}
\label{sec:bell-state}
%%%%%%%%%%

The measurement of number-sum Bell states plays the central role
in the present scheme for number-state manipulation.
The number-sum Bell states are given generally as
%%%%%
\begin{equation}
| {\bf d} (N,m) \rangle
= \sum_{k=0}^N d_k (N,m) | N - k \rangle_1 | k \rangle_2
\label{Bell-state}
\end{equation}
%%%%%
for $ m = 0 , 1 , \ldots , N $, forming an orthonormal set,
%%%%%
\begin{eqnarray}
\langle {\bf d} ( N^\prime , m^\prime ) | {\bf d} (N,m) \rangle
&=& \delta_{N^\prime N} {\bf d} ( N , m^\prime ) \cdot {\bf d} (N,m)
\nonumber \\
&=& \delta_{N^\prime N} \delta_{m^\prime m} .
\end{eqnarray}
%%%%%
The inner product of complex vectors is henceforth represented by
%%%%%
\begin{equation}
{\bf u} \cdot {\bf v} = \sum_{k=0}^N u_k^* v_k .
\end{equation}
%%%%%
The generic states in the two-mode Fock space
$ \{ | n_1 \rangle_1 | n_2 \rangle_2 \} $ are expanded
in terms of these Bell states as
%%%%%
\begin{eqnarray}
| \psi_{(2)} \rangle
&=&
\sum_{N = 0}^\infty \sum_{k = 0}^N
c_k (N) | N - k \rangle_1 | k \rangle_2
\nonumber \\
&=& \sum_{N = 0}^\infty \sum_{m = 0}^N
c_{\bf d} (N,m) | {\bf d} (N,m) \rangle ,
\end{eqnarray}
%%%%%
where
%%%%%
\begin{equation}
c_{\bf d} (N,m) = {\bf d} (N,m) \cdot {\bf c} (N)
= \sum_{k=0}^N d_k^* (N,m) c_k (N) .
\end{equation}
%%%%%

Specifically, we consider the number-phase Bell states
\cite{L-S,KY-2002,KY-2003,ZPM-2003,ZPM-2004},
%%%%%
\begin{equation}
| \phi_- (N,m) \rangle
= \sum_{k=0}^N \frac{[ ( \omega_{N+1}^* )^m ]^k}{\sqrt{N+1}}
| N - k \rangle_1 | k \rangle_2
\label{Bell-np1}
\end{equation}
%%%%%
with
%%%%%
\begin{equation}
d_k (N,m) = \frac{1}{\sqrt{N+1}} [ ( \omega_{N+1}^* )^m ]^k ,
\end{equation}
%%%%%
where the $ ( N+1 ) $-root to generate a $ {\rm Z}_{N+1} $ is given by
%%%%%
\begin{equation}
\omega_{N+1} \equiv \exp \left[ i 2 \pi /(N+1) \right] , \
( \omega_{N+1} )^{N+1} = 1 .
\end{equation}
%%%%%
These Bell states in Eq. (\ref{Bell-np1}) are also expressed as
%%%%%
\begin{equation}
| \phi_- (N,m) \rangle = \sum_{m^\prime = 0}^N
\frac{[ ( \omega_{N+1}^* )^{m^\prime + m} ]^N}{\sqrt{N+1}}
| \phi^{(N)}_{m^\prime + m} \rangle_1
| \phi^{(N)}_{m^\prime} \rangle_2
\label{Bell-np2}
\end{equation}
%%%%%
in terms of the phase states given by Pegg and Barnett \cite{P-B},
%%%%%
\begin{equation}
| \phi^{(N)}_m \rangle_p = \sum_{n=0}^N
\frac{[ ( \omega_{N+1} )^m ]^n}{\sqrt{N+1}} | n \rangle_p \ ( p = 1, 2 ) .
\label{phase-number}
\end{equation}
%%%%%

The Bell measurement of number-sum and phase-difference
is represented by the Hermitian operators,
%%%%%
\begin{eqnarray}
&& {\hat N}_+ \equiv {\hat N}_1 + {\hat N}_2 ,
\\
&& {\hat \Phi}_- \equiv \sum_{N=0}^\infty
[ {\hat \Phi}^{(N)}_1 - {\hat \Phi}^{(N)}_2 ] {\hat P}^{(N)} .
\end{eqnarray}
%%%%%
Here, $ {\hat N}_p $ ($ p = 1, 2 $) represent the number operators
of the respective modes, and $ {\hat \Phi}^{(N)}_p $ the phase operators
corresponding to the phase states in Eq. (\ref{phase-number}).
The projection operator $ {\hat P}^{(N)} $ extracts the states
in the subspace $ \{ | N - k \rangle_1 | k \rangle_2 \} $
with number-sum $ N $.
As seen clearly from Eqs. (\ref{Bell-np1}) and (\ref{Bell-np2}),
the Bell states $ | \phi_- (N,m) \rangle $
are the simultaneous eigenstates of number-sum and phase-difference:
%%%%%
\begin{eqnarray}
&& {\hat N}_+ | \phi_- (N,m) \rangle = N | \phi_- (N,m) \rangle ,
\label{egn-number}
\\
&& {\hat \Phi}_- | \phi_- (N,m) \rangle
= \phi_- (N,m) | \phi_- (N,m) \rangle ,
\label{egn-phase}
\end{eqnarray}
%%%%%
where the phase-difference eigenvalues are given by
%%%%%
\begin{equation}
\phi_- (N,m) = \frac{2 \pi}{N+1} m .
\end{equation}
%%%%%
Since $ [ {\hat \Phi}^{(N)}_1 - {\hat \Phi}^{(N)}_2 ] $ does not change
the number-sum $ N $, it commutes with $ {\hat P}^{(N)} $
as required for the Hermiticity of the entire phase-difference operator
$ {\hat \Phi}_- $.
These results clarify that in the subspace with number-sum $ N $
the phase-difference operator introduced by Luis and S\'anchez-Soto
\cite{L-S} indeed coincides
with the difference of the phase operators of the individual modes
given by Pegg and Barnett \cite{P-B},
while it is not separable in the entire two-mode Fock space.
It is also obvious from Eqs. (\ref{egn-number}) and (\ref{egn-phase})
that $ {\hat N}_+ $ and $ {\hat \Phi}_- $ are commutable:
%%%%%
\begin{equation}
[ {\hat N}_+ , {\hat \Phi}_- ] = 0 .
\end{equation}
%%%%%
Therefore, the joint measurement of number-sum and phase-difference
can be made in principle,
where the two-mode number states are projected
to the number-phase Bell states $ | \phi_- (N,m) \rangle $.

The number-phase Bell states in Eq. (\ref{Bell-np1}) may be generalized
by introducing a scaling parameter $ r $
\cite{KY-2002,KY-2003} as
%%%%%
\begin{equation}
| \phi_- (N,m,r) \rangle
= D(N,r) \sum_{k=0}^N r^k [ ( \omega_{N+1}^* )^m ]^k
| N - k \rangle_1 | k \rangle_2 ,
\end{equation}
%%%%%
where the normalization factor is given by
%%%%%
\begin{equation}
D(N,r) = \frac{{\sqrt{N+1}}( 1 - r^2 )}{1 - r^{2(N+1)}} .
\end{equation}
%%%%%
A two-mode squeezed vacuum state $ | \lambda \rangle $
with squeezing parameter $ \lambda < 1 $ may be used
as a primary resource of entanglement, which is given by
%%%%%
\begin{equation}
| \lambda \rangle = ( 1 - \lambda^2 )^{1/2}
\sum_{n=0}^\infty \lambda^n | n \rangle | n \rangle .
\end{equation}
%%%%%
Then, these generalized number-phase Bell states are actually generated
from a pair of two-mode squeezed vacuum states
$ | \lambda \rangle_{13} $ and $ | \lambda^\prime \rangle_{24} $
by making the number-phase Bell measurement:
%%%%%
\begin{equation}
| \lambda \rangle_{13} | \lambda^\prime \rangle_{24}
\stackrel{| \phi_- (N,-m) \rangle}{\Longrightarrow}
| \phi (N,m,r) \rangle ,
\end{equation}
%%%%%
where the scaling parameter $ r $ is given
by the ratio of the squeezing parameters,
%%%%%
\begin{equation}
r = \lambda^\prime / \lambda .
\end{equation}
%%%%%
Here, we have considered the relation
%%%%%
\begin{eqnarray}
| \lambda \rangle_{13} | \lambda^\prime \rangle_{24}
&=& ( 1 - \lambda^2 )^{1/2} ( 1 - \lambda^{\prime 2} )^{1/2}
\sum_{N=0}^\infty \frac{\lambda^N}{D(N,r)}
\nonumber
\\
& \times & 
\sum_{m=0}^N | \phi_- (N,-m) \rangle_{34} | \phi_- (N,m,r) \rangle_{12}
\label{1234-bell}
\end{eqnarray}
%%%%%
from the swapping $ (1,3)(2,4) \rightarrow (1,2)(3,4) $.

%%%%%%%%%%
\section{Practical Bell state detector}
\label{sec:bell-detector}
%%%%%%%%%%

\begin{figure}[t]
\centering
\scalebox{.38}{\includegraphics*[0cm,1.5cm][25cm,22.5cm]{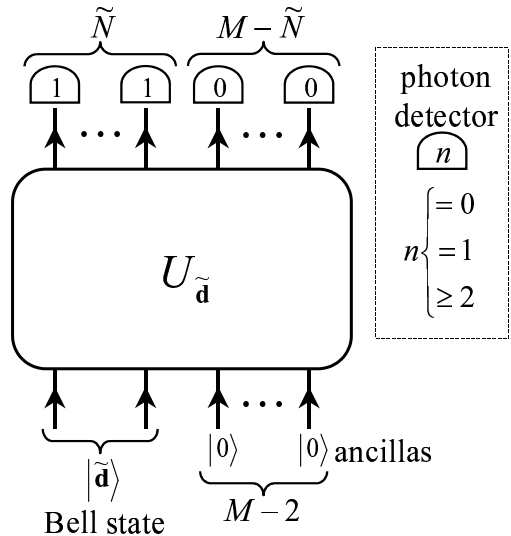}}
\caption{
A schematic diagram of the linear optical Bell state detector.
The input two-mode state,
which may contain the number-sum Bell state
$ | {\tilde{\bf d}} \rangle
\equiv | {\bf d} ({\tilde N},{\tilde m}) \rangle$
with number-sum $ {\tilde N} $,
enters the detector together with the vacuum states
of $ ( M - 2 ) $ ancilla modes.
A unitary transformation $ U_{\tilde{\bf d}} $ is made
through a set of beam splitters and phase shifters,
and the output state is detected to give conditionally
the specific photon count
$ {\bf n}_{(M)}^{\rm cnt} = ( 1 , \ldots , 1 , 0 , \ldots , 0 ) $
as the signal of $ | {\tilde{\bf d}} \rangle $.
The photon detectors need to resolve zero, one or more photons,
since two or more photons may enter some of the detectors
for the case of $ {\tilde N} \geq 2 $.
}
\label{bsdetector}
\end{figure}

We utilize a linear optical detector, say {\it Bell state detector},
to measure conditionally
a specific two-mode number-sum Bell state
as given in Eq. (\ref{Bell-state}).
Henceforth the Bell state to be detected is denoted simply by
%%%%%
\begin{equation}
| {\tilde{\bf d}} \rangle
\equiv | {\bf d} ({\tilde N},{\tilde m}) \rangle
\end{equation}
%%%%%
with the number-sum $ {\tilde N} $ and amplitude coefficients
%%%%%
\begin{equation}
{\tilde d}_k \equiv d_k ({\tilde N},{\tilde m}) .
\end{equation}
%%%%%
As shown schematically in Fig. \ref{bsdetector},
it is constructed as an $ M $-port system consisting of
(i) a set of beam splitters and phase shifters,
(ii) $ ( M - 2 ) $ auxiliary input modes (ancillas) with vacuum states,
and (iii) zero-one-resolving photon detectors for the output modes,
though imperfect practically.
This method is based on the idea of photon chopping
\cite{chopping}.
The Bell state detectors of $ | \phi_- ({\tilde N},{\tilde m}) \rangle $
for $ {\tilde N} = 1 $ and $ 2 $
are considered in Refs. \cite{KY-2002,KY-2003},
and then a method for general $ {\tilde N} $ is presented
in Refs. \cite{ZPM-2003,ZPM-2004}.
The photon detectors need to resolve zero, one or more photons,
since two or more photons may enter some of the detectors
for the case of $ {\tilde N} \geq 2 $.
%%%%%
%%%%%

The operation of the set of beam splitters and phase shifters
is given by a unitary transformation between the input modes $ a_i $
and the output modes $ b_j $ (in Heisenberg picture)
\cite{Reck-Zeilinger}:
%%%%%
\begin{equation}
a_i = {\cal U}_{\tilde{\bf d}} b_i {\cal U}_{\tilde{\bf d}}^\dagger
= U_{\tilde{\bf d}ij} b_j , \
a_i^\dagger
= {\cal U}_{\tilde{\bf d}} b_i^\dagger {\cal U}_{\tilde{\bf d}}^\dagger
= b_j^\dagger U_{\tilde{\bf d}ji}^\dagger ,
\label{Ud}
\end{equation}
%%%%%
where $ i , j = 1 , 2 , \ldots , M $,
and $ U_{\tilde{\bf d}} $ is an $ M \times M $ unitary matrix.
The two-mode input state $ | \psi_{(2)} \rangle $
and vacuum state $ | {\bf 0} \rangle_{\bf a} $ of $ ( M - 2 ) $ ancillas
are transformed to certain output state through the optical set
(in Schr\"odinger picture),
which may be expanded in terms of the number-states
of the output $ M $ modes,
%%%%%
\begin{eqnarray}
| {\bf n}_{(M)} \rangle
\equiv | n_1 \rangle_1 | n_2 \rangle_2 \cdots | n_M \rangle_M
\end{eqnarray}
%%%%%
with number distribution
%%%%%
\begin{eqnarray}
{\bf n}_{(M)} \equiv ( n_1 , n_2 , \ldots , n_M ) .
\end{eqnarray}
%%%%%
The parameters of the optical set are chosen
so that this unitary transformation is given as
%%%%%
\begin{equation}
{\cal U}_{\tilde{\bf d}} | \psi_{(2)} \rangle | {\bf 0} \rangle_{\bf a}
= g_{\tilde {\bf d}} \langle {\tilde{\bf d}} | \psi_{(2)} \rangle
| {\bf n}_{(M)}^{\rm cnt} \rangle
+ | \Psi \bot {\bf n}_{(M)}^{\rm cnt} \rangle ,
\label{U-psi2}
\end{equation}
%%%%%
where $ | \Psi \bot {\bf n}_{(M)}^{\rm cnt} \rangle $
is a certain state orthogonal to $ | {\bf n}_{(M)}^{\rm cnt} \rangle $.
That is, only if the input state $ | \psi_{(2)} \rangle $
contains the Bell state $ | {\tilde{\bf d}} \rangle $ to be detected,
the output state has the component of the specific number distribution,
%%%%%
\begin{equation}
{\bf n}_{(M)}^{\rm cnt}
= ( \overbrace{1 , \ldots , 1}^{\tilde N} ,
\overbrace{0 , \ldots , 0}^{M-{\tilde N}} ) .
\label{photon-count}
\end{equation}
%%%%%
Then, by using the ideal zero-one-resolving photon detectors,
the Bell state $ | {\tilde{\bf d}} \rangle $
is detected conditionally in $ | \psi_{(2)} \rangle $,
when the photon counting result of $ {\bf n}_{(M)}^{\rm cnt} $
is obtained with success probability
%%%%%
\begin{equation}
P_{\tilde{\bf d}}^{\rm ideal} [ | \psi_{(2)} \rangle ]
= | g_{\tilde{\bf d}} |^2
| \langle {\tilde{\bf d}} | \psi_{(2)} \rangle |^2 .
\end{equation}
%%%%%
Practically, we use imperfect zero-one-resolving photon detectors
described by the POM's $ \Pi (0) $ and $ \Pi (1) $.
The POM of photon detector for the $ N $ photon count
is given with quantum efficiency $ \eta $ and mean dark count $ \nu $
by
%%%%%
\begin{eqnarray}
\Pi (N)
& = & \sum_{m=0}^N {\rm e}^{- \nu} \frac{\nu^{N - m}}{( N - m )!}
\nonumber \\
& \times &
\sum_{n=m}^\infty {}_n C_m \eta^m ( 1 - \eta )^{n - m}
| n \rangle \langle n | ,
\label{pd-POM}
\end{eqnarray}
%%%%%
where $ {}_n C_m $ is the binomial coefficient
\cite{Barnett-Phillips-Pegg}.

The two-mode input state $ \rho_{(2)} $
combined with the ancilla-mode $ \rho_{\bf a} $
is transformed by the optical set as
%%%%%
\begin{equation}
\rho_{(2)} \otimes \rho_{\bf a}
\rightarrow {\cal U}_{\tilde{\bf d}} \rho_{(2)} \otimes \rho_{\bf a}
{\cal U}_{\tilde{\bf d}}^\dagger ,
\end{equation}
%%%%%
where
%%%%%
\begin{eqnarray}
\rho_{\bf a} = | {\bf 0} \rangle_{\bf a} {}_{\bf a} \langle {\bf 0} |
\equiv \mathop{\otimes}_{i=3}^{M} ( | 0 \rangle \langle 0 | )_i .
\end{eqnarray}
%%%%%
Then, the probability to obtain the photon count
of Eq. (\ref{photon-count}) for the two-mode $ \rho_{(2)} $
is given by
%%%%%
\begin{eqnarray}
P_{\tilde{\bf d}} [ \rho_{(2)} ]
&=& {\rm Tr} [ \Pi_{\rm PD} {\cal U}_{\tilde{\bf d}}
\rho_{(2)} \otimes \rho_{\bf a} {\cal U}_{\tilde{\bf d}}^\dagger ]
\nonumber \\
& \equiv & {\rm Tr} [ \Gamma_{\tilde{\bf d}} \rho_{(2)} ] .
\label{P-rho}
\end{eqnarray}
%%%%%
The POM of this Bell state detector is given by
%%%%%
\begin{equation}
\Gamma_{\tilde{\bf d}}
= {}_{\bf a} \langle {\bf 0} |  {\cal U}_{\tilde{\bf d}}^\dagger
\Pi_{\rm PD} {\cal U}_{\tilde{\bf d}} | {\bf 0} \rangle_{\bf a}
\end{equation}
%%%%%
with the POM of the photon detector set
%%%%%
\begin{eqnarray}
&& \Pi_{\rm PD} = \mathop{\otimes}_{i=1}^{\tilde N} \Pi (1)_i
\mathop{\otimes}_{i={\tilde N}+1}^{M} \Pi (0)_i .
\end{eqnarray}
%%%%%
It may be expressed as
%%%%%
\begin{equation}
\Gamma_{\tilde{\bf d}}
= \sum_{N=0}^\infty \sum_{k=0}^N \sum_{k^\prime = 0}^N
K^{\tilde{\bf d}}_{k^\prime k} (N)
| ( N , k^\prime ) \rangle \langle ( N , k ) |
\label{POM-Bell}
\end{equation}
%%%%%
in terms of the basis states of number-sum $ N $
%%%%%
\begin{equation}
| ( N , k ) \rangle \equiv | N - k \rangle_1 | k \rangle_2
\end{equation}
%%%%%
with
%%%%%
\begin{equation}
K^{\tilde{\bf d}}_{k^\prime k} (N) \delta_{N^\prime N}
= \langle ( N^\prime , k^\prime ) | \Gamma_{\tilde{\bf d}}
| ( N , k ) \rangle .
\end{equation}
%%%%%
Here, it should be remarked that
the matrix elements of $ \Gamma_{\tilde{\bf d}} $
between the states with different values of number-sum are zero,
since $ {\cal U}_{\tilde{\bf d}} $ and $ \Pi_{\rm PD} $
conserve the total photon number.

Specifically, for the basis state $ | (N,k) \rangle $
we obtain the output state as
%%%%%
\begin{equation}
{\cal U}_{\tilde{\bf d}} | (N,k) \rangle | {\bf 0} \rangle_{\bf a}
= \sum_{N_\Sigma [ {\bf n}_{(M)} ] = N}
B^{\tilde{\bf d}}_k [{\bf n}_{(M)} ] | {\bf n}_{(M)} \rangle .
\label{UdNk}
\end{equation}
%%%%%
Here, the sum is taken over the distributions
$ {\bf n}_{(M)} $ with number-sum $ N $,
since the unitary transformation $ {\cal U}_{\tilde{\bf d}} $
conserves the total photon number
%%%%%
\begin{equation}
N_\Sigma [ {\bf n}_{(M)} ] \equiv \sum_{i=1}^{M} n_i = N .
\end{equation}
%%%%%
The basis states with number-sum $ N $ are given by
%%%%%
\begin{equation}
| (N,k) \rangle
= \frac{1}{{\sqrt{(N - k)!}}{\sqrt{k !}}}
( a_1^\dagger )^{N - k} ( a_2^\dagger )^k
| 0 \rangle_1 | 0 \rangle_2 .
\end{equation}
%%%%%
By using Eq. (\ref{Ud}) we obtain
%%%%%
\begin{equation}
( a_1^\dagger )^{N - k} ( a_2^\dagger )^k
= \sum_{{\bf j}_{(N)}}
W^{\tilde{\bf d}}_k [ {\bf j}_{(N)} ]
\mathop{\otimes}_{l=1}^N b_{j_l}^\dagger ,
\end{equation}
%%%%%
where $ {\bf j}_{(N)} \equiv ( j_1 , j_2 , \ldots , j_N ) $,
$ 1 \leq j_l \leq M $,
and
%%%%%
\begin{equation}
W^{\tilde{\bf d}}_k [ {{\bf j}_{(N)}} ]
= U_{{\tilde{\bf d}}1j_1}^* \cdots U_{{\tilde{\bf d}}1j_{N - k}}^*
U_{{\tilde{\bf d}}2j_{N - k + 1}}^* \cdots U_{{\tilde{\bf d}}2j_N}^* .
\end{equation}
%%%%%
Then, we calculate the coefficients
for the output state $ | {\bf n}_{(M)} \rangle $ in Eq. (\ref{UdNk}) as
%%%%%
\begin{equation}
B^{\tilde{\bf d}}_k [ {\bf n}_{(M)} ]
= \frac{\sqrt{{\bf n}_{(M)}!}}{{\sqrt{(N - k)!}}{\sqrt{k !}}}
\sum_{{\bf j}_{(N)} \to {\bf n}_{(M)}}
W^{\tilde{\bf d}}_k [ {\bf j}_{(N)} ] ,
\end{equation}
%%%%%
where $ {\bf n}_{(M)}! \equiv n_1 ! n_2 ! \cdots n_M ! $,
and the sum is taken over all the sets of indices
$ {\bf j}_{(N)} $ that provide the photon-number distribution
$ {\bf n}_{(M)} $.

Given the the coefficients $ B^{\tilde{\bf d}}_k [ {\bf n}_{(M)} ] $
in Eq. (\ref{UdNk}), we obtain the matrix elements of Bell measurement POM
$ \Gamma_{\tilde{\bf d}} $ in Eq. (\ref{POM-Bell}) as
%%%%%
\begin{equation}
K^{\tilde{\bf d}}_{k^\prime k} (N)
= \sum_{N_\Sigma [ {\bf n}_{(M)} ] = N}
B^{{\tilde{\bf d}} *}_{k^\prime} [ {\bf n}_{(M)} ]
B^{\tilde{\bf d}}_k [ {\bf n}_{(M)} ]
P_{\rm PD} [ | {\bf n}_{(M)} \rangle ] .
\label{K-Bell}
\end{equation}
%%%%%
Here, we have considered the relation from the photon-number conserving
nature of $ \Pi_{\rm PD} $,
%%%%%
\begin{equation}
\langle {\bf n}_{(M)}^\prime | \Pi_{\rm PD} | {\bf n}_{(M)} \rangle
= \delta_{{\bf n}_{(M)}^\prime {\bf n}_{(M)}}
P_{\rm PD} [ | {\bf n}_{(M)} \rangle ] .
\end{equation}
%%%%%
The probability that the state $ | {\bf n}_{(M)} \rangle $
results in the photon count $ {\bf n}_{(M)}^{\rm cnt} $
is given by
%%%%%
\begin{eqnarray}
P_{\rm PD} [ | {\bf n}_{(M)} \rangle ]
&=& \langle {\bf n}_{(M)} | \Pi_{\rm PD} | {\bf n}_{(M)} \rangle
\nonumber \\
&=& \prod_{i=1}^{\tilde N} P_{1 \gamma} ( n_i )
\prod_{i={\tilde N}+1}^{M} P_{0 \gamma} ( n_i ) ,
\end{eqnarray}
%%%%%
where
%%%%%
\begin{eqnarray}
&& P_{0 \gamma} (n) = \langle n | \Pi (0) | n \rangle
= {\rm e}^{- \nu} {\delta \eta}^n ,
\\
&& P_{1 \gamma} (n) = \langle n | \Pi (1) | n \rangle
= {\rm e}^{- \nu} {\delta \eta}^{n-1}
[ n ( 1 - {\delta \eta} ) + \nu {\delta \eta} ]
\nonumber \\
{}
\end{eqnarray}
%%%%%
with
%%%%%
\begin{equation}
\delta \eta \equiv 1 - \eta .
\end{equation}
%%%%%

The output state $ | {\bf n}_{(M)}^{\rm cnt} \rangle $, in particular,
to indicate the desired Bell state $ | {\tilde{\bf d}} \rangle $
is faithfully counted as $ {\bf n}_{(M)}^{\rm cnt} $
with probability 
%%%%%
\begin{equation}
P_{\rm PD} [ | {\bf n}_{(M)}^{\rm cnt} \rangle ]
= {\rm e}^{- M \nu} [ 1 - {\delta \eta} + \nu {\delta \eta} ]^{\tilde N} .
\end{equation}
%%%%%
(Henceforth we assume for simplicity that all the photon detectors
have the common $ \eta $ and $ \nu $.)
The probability for the generic output state $ | {\bf n}_{(M)} \rangle $
to give the expected photon count $ {\bf n}_{(M)}^{\rm cnt} $
is also evaluated as
%%%%%
\begin{equation}
P_{\rm PD} [ | {\bf n}_{(M)} \rangle ]
= {\rm e}^{- M \nu} \sum_{(a,b)}
r_{\rm PD}^{(a,b)} [ {\bf n}_{(M)} ]
{\delta \eta}^a \nu^b ( 1 - {\delta \eta} )^{{\tilde N} - b}
\label{rPD}
\end{equation}
%%%%%
with certain coefficients $ r_{\rm PD}^{(a,b)} [ {\bf n}_{(M)} ] $,
where the extra factors $ ( 1 - {\delta \eta} )^{{\tilde N} - b} $
come from $ P_{1 \gamma} ( n_i ) $ ($ 1 \leq i \leq {\tilde N} $).
The non-negative powers $ a $ and $ b $
in the expansion of Eq. (\ref{rPD})
represent the discounts and overcounts of photons, respectively,
which satisfy the relation
%%%%%
\begin{equation}
a - b = N - {\tilde N}
\label{a-b}
\end{equation}
%%%%%
in the range of $ {\rm max} [ 0 , N - {\tilde N} ] \leq a \leq N $
and $ {\rm max} [ 0 , {\tilde N} - N ] \leq b \leq {\tilde N} $.
For $ N < {\tilde N} $ the deficit of photons should be supplied
by the dark counts,
while for $ N > {\tilde N} $
the excess of photons should be discarded with $ \eta < 1 $.
By considering Eq. (\ref{a-b}),
the leading dependence of $ P_{\rm PD} [ | {\bf n}_{(M)} \rangle ] $
on $ {\delta \eta} < 1 $ and $ \nu < 1 $
is found for the output states other
than $ | {\bf n}_{(M)}^{\rm cnt} \rangle $ as
%%%%%
\begin{equation}
P_{\rm PD} [ | {\bf n}_{(M)} \rangle
\not= | {\bf n}_{(M)}^{\rm cnt} \rangle ]
\sim \left\{ \begin{array}{ll}
\nu^{{\tilde N} - N} & ( N < {\tilde N} ) \\
{\delta \eta} \nu & ( N = {\tilde N} ) \\
{\delta \eta}^{N - {\tilde N}} & ( N > {\tilde N} )
\end{array} \right. .
\end{equation}
%%%%%
It may be reasonably assumed for feasible photon detectors
that the dark count $ \nu $ is considerably smaller
than the inefficiency $ {\delta \eta} $,
e.g., $ \nu \sim 10^{-4} $ and $ {\delta \eta} \sim 0.1 $,
as will be explained in Sec. \ref{sec:analysis}.
Then, the leading error $ \sim {\delta \eta} $ of the Bell state detector
is provided by the states $ | {\bf n}_{(M)} \rangle $
with the total photon number $ N = {\tilde N} + 1 $.

When the desired Bell state $ | {\tilde{\bf d}} \rangle $
is measured by this Bell state detector,
the probability to obtain the expected photon count
$ {\bf n}_{(M)}^{\rm cnt} $ is given
with Eqs. (\ref{P-rho}), (\ref{POM-Bell}) and (\ref{K-Bell}) as
%%%%%
\begin{eqnarray}
P_{\tilde{\bf d}} [ | {\tilde{\bf d}} \rangle ]
&=& {\rm Tr} [ \Gamma_{\tilde{\bf d}}
| {\tilde{\bf d}} \rangle \langle {\tilde{\bf d}} | ]
\nonumber \\
&=& \sum_{N_\Sigma [ {\bf n}_{(M)} ] = {\tilde N}}
P_{\rm PD} [ | {\bf n}_{(M)} \rangle ]
\left| {\bf B}^{\tilde{\bf d}} [ {\bf n}_{(M)} ] \cdot {\tilde {\bf d}}
\right|^2 .
\nonumber \\
\label{P-Bell}
\end{eqnarray}
%%%%%
In this practical Bell measurement,
the other states orthogonal to $ | {\tilde{\bf d}} \rangle $
may be miscounted as $ | {\tilde{\bf d}} \rangle $
with nonzero probabilities.
Only if we can use the ideal Bell state detector,
the Bell state is measured faithfully as
%%%%%
\begin{equation}
P_{\tilde{\bf d}}^{\rm ideal} [ | {\bf d} (N,m) \rangle ]
= {\bar P}_{\tilde{\bf d}} [ | {\tilde{\bf d}} \rangle ]
\delta_{N{\tilde N}} \delta_{m{\tilde m}} .
\end{equation}
%%%%%
That is, the desired Bell state $ | {\tilde{\bf d}} \rangle $
is measured with the success probability
$ {\bar P}_{\tilde{\bf d}} [ | {\tilde{\bf d}} \rangle ] $,
while the other orthogonal states are not detected.
By considering Eq. (\ref{U-psi2})
with $ P_{\rm PD}^{\rm ideal} [ | {\bf n}_{(M)}^{\rm cnt} \rangle ] = 1 $,
the success probability in the ideal case is evaluated as
%%%%%
\begin{equation}
{\bar P}_{\tilde{\bf d}} [ | {\tilde{\bf d}} \rangle ]
= | g_{\tilde{\bf d}} |^2
= \left| {\bf B}^{\tilde{\bf d}} [ {\bf n}_{(M)}^{\rm cnt} ]
\cdot {\tilde {\bf d}} \right|^2 .
\label{P-Bell-ideal}
\end{equation}
%%%%%
On the other hand, from the completeness of number-state Fock space
the sum of the probabilities
for the orthonormal basis states
$ | {\bf d} (N,m) \rangle \equiv | {\tilde{\bf d}}_\bot \rangle $
other than $ | {\tilde{\bf d}} \rangle $
to be miscounted as $ | {\tilde{\bf d}} \rangle $ is given by
%%%%%
\begin{equation}
\sum_{| {\tilde{\bf d}}_\bot \rangle}
{\rm Tr} [ \Gamma_{\tilde{\bf d}}
[ | {\tilde{\bf d}}_\bot \rangle \langle {\tilde{\bf d}}_\bot | ]
= {\rm Tr} [ \Gamma_{\tilde{\bf d}} ]
- {\rm Tr} [ \Gamma_{\tilde{\bf d}}
[ | {\tilde{\bf d}} \rangle \langle {\tilde{\bf d}} | ] ,
\label{miscount}
\end{equation}
%%%%%
where
%%%%%
\begin{eqnarray}
{\bf 1} &=&
| {\tilde{\bf d}} \rangle \langle {\tilde{\bf d}} |
+ \sum_{| {\tilde{\bf d}}_\bot \rangle}
| {\tilde{\bf d}}_\bot \rangle \langle {\tilde{\bf d}}_\bot | ,
\\
{\rm Tr} [ \Gamma_{\tilde{\bf d}} ]
&=& \sum_{{\bf n}_{(M)}} P_{\rm PD} [ | {\bf n}_{(M)} \rangle ]
\left| {\bf B}^{\tilde{\bf d}} [ {\bf n}_{(M)} ] \right|^2 .
\label{TrGd}
\end{eqnarray}
%%%%%
Then, the {\it confidence} of this practical Bell state detector
may be defined by
%%%%%
\begin{eqnarray}
{\cal C}_{\tilde{\bf d}}
&=& \frac{{\rm Tr} [ \Gamma_{\tilde{\bf d}}
| {\tilde{\bf d}} \rangle \langle {\tilde{\bf d}} | ]}
{{\rm Tr} [ \Gamma_{\tilde{\bf d}} ]} \leq 1 .
\label{confidence}
\end{eqnarray}
%%%%%
In particular, $ {\cal C}_{\tilde{\bf d}} = 1 $
only for the Bell state detector with ideal optical devices.
We evaluate the confidence in Eq. (\ref{confidence})
with Eqs. (\ref{P-Bell}) and (\ref{TrGd})
for the practical Bell state detector as
%%%%%
\begin{eqnarray}
{\cal C}_{\tilde{\bf d}}
& \equiv & 1 - \Delta {\cal C}_{\tilde{\bf d}}
\nonumber \\
&=& 1 - \sum_{(a,b) \not= (0,0)}
q_{\tilde{\bf d}}^{(a,b)}
{\delta \eta}^a \nu^b
\end{eqnarray}
%%%%%
in the expansion with respect to $ \delta \eta $ and $ \nu $.

%%%%%%%%%%
\section{Number-state manipulation}
\label{sec:manipulation}
%%%%%%%%%%

We now investigate the number-state manipulation
via teleportation with number-sum Bell measurement.
The input state (normalized) may be prepared in $ K $ optical modes as
%%%%%
\begin{equation}
| \psi_{\rm in} \rangle = \sum_{{\bf n}_{(K)}}
c^{\rm in}_{{\bf n}_{(K)}} | {{\bf n}_{(K)}} \rangle ,
\end{equation}
%%%%%
where
%%%%%
\begin{equation}
| {{\bf n}_{(K)}} \rangle
\equiv | n_1 \rangle_1 | n_2 \rangle_2
\cdots | n_K \rangle_K
\equiv | n_1 \rangle_1 | {{\bf n}_{(K-1)}} \rangle .
\end{equation}
%%%%%
We here consider specifically a class of two-mode EPR resources
(normalized) as
%%%%%
\begin{equation}
| {\rm EPR} \rangle = \sum_{l=0}^\infty
E_l | l \rangle_0 | s (l) \rangle_{-1}
\equiv \sum_{l=0}^\infty \sum_{l^\prime = 0}^\infty
E_{l^\prime l} | l \rangle_0 | l^\prime \rangle_{-1}
\end{equation}
%%%%%
with the amplitude distribution matrix
%%%%%
\begin{equation}
E_{l^\prime l} = \delta_{l^\prime s (l)} E_l .
\end{equation}
%%%%%
The permutation of number-states between the two modes
is given by
%%%%%
\begin{equation}
s ( l_1 ) \not= s ( l_2 ) \leftrightarrow l_1 \not= l_2 .
\end{equation}
%%%%%
In particular, for the number-difference 0 resource
and the number-sum $ N $ resource, respectively,
%%%%%
\begin{equation}
s(l) = \left\{ \begin{array}{ll}
l & ({\mbox{number-difference 0}}) \\
N - l & ({\mbox{number-sum $ N $}}) \end{array} \right. .
\end{equation}
%%%%%
The input state is then manipulated by making a Bell measurement
with an EPR resource.
We here consider the one-mode manipulation
with the measurement of $ | {\tilde{\bf d}} \rangle $.
The multimode manipulation may further be performed
by applying these sorts of one-mode manipulations
to some modes of the input state.

The Bell measurement is made on the 0-1 mode of the combined state
($ n_1 \equiv k $, $ l \equiv N - k $)
%%%%%
\begin{eqnarray}
| \psi_{\rm in} \rangle | {\rm EPR} \rangle
&=& \sum_{{\bf n}_{(K-1)}}
\sum_{N=0}^\infty \sum_{k=0}^N
c^{\rm in}_{k{\bf n}_{(K-1)}} E_{N-k} | ( N , k ) \rangle
\nonumber \\
& \times &
| s(N-k) \rangle_{-1} | {{\bf n}_{(K-1)}} \rangle .
\end{eqnarray}
%%%%%
Then, we obtain the output state as
%%%%%
\begin{eqnarray}
\rho_{\rm out}
&=& {\rm Tr} [ \Gamma_{\tilde{\bf d}}
\rho_{\rm in} \otimes \rho_{\rm EPR} ]
\nonumber \\
&=& \sum_{{\bf n}_{(M)}} P_{\rm PD} [ | {\bf n}_{(M)} \rangle ]
| \psi_{\rm out} [ {\bf n}_{(M)} ] \rangle
\langle \psi_{\rm out} [ {\bf n}_{(M)} ] |
\nonumber \\
&{}&
\end{eqnarray}
%%%%%
for $ \rho_{\rm in} = | \psi_{\rm in} \rangle \langle \psi_{\rm in} | $
and $ \rho_{\rm EPR} = | {\rm EPR} \rangle \langle {\rm EPR} | $,
where $ | s(N-k) \rangle_{-1} | {{\bf n}_{(K-1)}} \rangle
\equiv | {\bf n}_{(K)} \rangle $
by redenoting
%%%%%
\begin{equation}
| s(N-k) \rangle_{-1} \equiv | n_1 \rangle_1 .
\end{equation}
%%%%%
(The output state $ \rho_{\rm out} $ will be properly normalized later
in defining the fidelity.)
The output states associated with $ | {\bf n}_{(M)} \rangle $,
which may not be orthogonal each other, are given by
%%%%%
\begin{equation}
| \psi_{\rm out} [ {\bf n}_{(M)} ] \rangle
= \sum_{{\bf n}_{(K)}} c^{\rm out}_{{\bf n}_{(K)}} [ {{\bf n}_{(M)}} ]
| {\bf n}_{(K)} \rangle
\end{equation}
%%%%%
with the amplitudes
%%%%%
\begin{equation}
c^{\rm out}_{{\bf n}_{(K)}} [ {\bf n}_{(M)} ]
= \left. E_{N-k} B^{\tilde{\bf d}}_k [ {\bf n}_{(M)} ]
c^{\rm in}_{k{\bf n}_{(K-1)}}
\right|^{N = N_\Sigma [ {\bf n}_{(M)} ]}_{n_1 = s(N-k)} ,
\label{c-out-nK}
\end{equation}
%%%%%
where $ k $ is specified by $ n_1 = s(N-k) $
in terms of $ n_1 $ and $ N $.
It is straightforward to extend these formulas
generally for the mixed states of $ \rho_{\rm in} $ and $ \rho_{\rm EPR} $
with the output states as
$ | \psi_{\rm out} [ {\bf n}_{(M)} ] \rangle
\langle \psi_{\rm out} [ {\bf n}_{(M)} ] |
\rightarrow  \rho_{\rm out} [ {\bf n}_{(M)} ] $.

This teleportation-based manipulation may be viewed
as a linear transformation of the input state:
%%%%%
\begin{eqnarray}
\rho_{\rm out}
&=& {\cal T}_{{\tilde{\bf d}}{\bf E}} \rho_{\rm in}
{\cal T}_{{\tilde{\bf d}}{\bf E}}^\dagger
\nonumber \\
& \equiv & \sum_{{\bf n}_{(M)}} P_{\rm PD} [ | {\bf n}_{(M)} \rangle ]
\left( {\cal T}_{{\tilde{\bf d}}{\bf E}} [ {\bf n}_{(M)} ]
\rho_{\rm in}
{\cal T}_{{\tilde{\bf d}}{\bf E}}^\dagger [ {\bf n}_{(M)} ] \right) .
\nonumber \\
\end{eqnarray}
%%%%%
The amplitudes are accordingly transformed as
%%%%%
\begin{equation}
{\bf c}^{\rm out} [ {\bf n}_{(M)} ]
= {\bf T}^{{\tilde{\bf d}}{\bf E}} [ {\bf n}_{(M)} ] {\bf c}^{\rm in}
\label{c-out}
\end{equation}
%%%%%
or
%%%%%
\begin{equation}
c^{\rm out}_{{\bf n}_{(K)}} [ {\bf n}_{(M)} ]
= \sum_{k=0}^N T^{{\tilde{\bf d}}{\bf E}}_{n_1 k} [ {\bf n}_{(M)} ]
c^{\rm in}_{k{\bf n}_{(K-1)}} .
\end{equation}
%%%%%
As seen from Eq. (\ref{c-out-nK}), the transformation matrix
$ {\bf T}^{{\tilde{\bf d}}{\bf E}} [ {\bf n}_{(M)} ] $
is composed of that given by the Bell state detector,
$ {\bf B}^{\tilde{\bf d}} [ {\bf n}_{(M)} ] $,
the reversal ($ 0 , \ldots , N \rightarrow N , \ldots , 0 $)
with $ N = N_\Sigma [ {\bf n}_{(M)} ] $, $ {\bf R}_N $,
and the EPR resource, $ {\bf E} $:
%%%%%
\begin{equation}
{\bf T}^{{\tilde{\bf d}}{\bf E}} [ {\bf n}_{(M)} ]
= {\bf E} {\bf R}_N {\bf B}^{\tilde{\bf d}} [ {\bf n}_{(M)} ] ,
\label{TdE}
\end{equation}
%%%%%
where
%%%%%
\begin{equation}
( {\bf B}^{\tilde{\bf d}} [ {\bf n}_{(M)} ] )_{k k^\prime}
= \delta_{k k^\prime} B^{\tilde{\bf d}}_k [ {\bf n}_{(M)} ]
\theta ( N - k )
\end{equation}
%%%%%
with
%%%%%
\begin{equation}
\theta ( N - k ) = \left\{ \begin{array}{ll}
1 & ( 0 \leq k \leq N ) \\ 0 & ( k > N ) \end{array} \right. .
\end{equation}
%%%%%
We may further consider multiple of manipulations of this sort
\cite{KY-2003} as
%%%%%
\begin{equation}
{\cal T}^{{\tilde{\bf d}}{\bf E}(L)} \cdots
{\cal T}^{{\tilde{\bf d}}{\bf E}(2)}
{\cal T}^{{\tilde{\bf d}}{\bf E}(1)} .
\end{equation}
%%%%%

The desired manipulation of input state with the EPR resource
is obtained by using the ideal Bell state detector
of $ | {\tilde{\bf d}} \rangle $ as
%%%%%
\begin{equation}
\rho_{\rm out} ( \eta = 1 , \nu = 0 )
= | \psi_{\rm out}^{{\tilde{\bf d}}{\bf E}} \rangle
\langle \psi_{\rm out}^{{\tilde{\bf d}}{\bf E}} | ,
\end{equation}
%%%%%
where
%%%%%
\begin{equation}
| \psi_{\rm out}^{{\tilde{\bf d}}{\bf E}} \rangle
= | \psi_{\rm out} [ {\bf n}_{(M)} = {\bf n}_{(M)}^{\rm cnt} ] \rangle
\end{equation}
%%%%%
with
%%%%%
\begin{equation}
c^{\rm out}_{{\bf n}_{(K)}} [ {\bf n}_{(M)}^{\rm cnt} ]
= \sum_{k=0}^N
T^{{\tilde{\bf d}}{\bf E}}_{n_1 k} [ {\bf n}_{(M)}^{\rm cnt} ]
c^{\rm in}_{k{\bf n}_{(K-1)}} .
\end{equation}
%%%%%
Here, only the number-state $ | {\bf n}_{(M)}^{\rm cnt} \rangle $
is detected faithfully as the photon count $ {\bf n}_{(M)}^{\rm cnt} $
in the output $ M $ ports.
The {\it fidelity} is used to evaluate the quality of manipulation
with the practical experimental setup, which is given by
%%%%%
\begin{equation}
F [ | \psi_{\rm out}^{{\tilde{\bf d}}{\bf E}} \rangle ]
= \frac{{\rm Tr} [ \rho_{\rm out}
| \psi_{\rm out}^{{\tilde{\bf d}}{\bf E}}  \rangle
\langle \psi_{\rm out}^{{\tilde{\bf d}}{\bf E}}  | ]}
{{\rm Tr} [ \rho_{\rm out} ]
{\rm Tr} [ | \psi_{\rm out}^{{\tilde{\bf d}}{\bf E}} \rangle
\langle \psi_{\rm out}^{{\tilde{\bf d}}{\bf E}} | ]}
\leq 1 ,
\end{equation}
%%%%%
where the denominator of the right side
provides the normalization factors of $ \rho_{\rm out} $
and $ | \psi_{\rm out}^{{\tilde{\bf d}}{\bf E}} \rangle
\langle \psi_{\rm out}^{{\tilde{\bf d}}{\bf E}} | $.
The relevant quantities are calculated by
%%%%%
\begin{eqnarray}
{\rm Tr} [ \rho_{\rm out} ]
&=& \sum_{{\bf n}_{(M)}} P_{\rm PD} [ | {\bf n}_{(M)} \rangle ]
\nonumber \\
& \times &
{\bf c}^{\rm out} [ {\bf n}_{(M)} ]
\cdot {\bf c}^{\rm out} [ {\bf n}_{(M)} ] ,
\\
{\rm Tr} [ | \psi_{\rm out}^{{\tilde{\bf d}}{\bf E}} \rangle
\langle \psi_{\rm out}^{{\tilde{\bf d}}{\bf E}} | ]
&=& {\bf c}^{\rm out} [ {\bf n}_{(M)}^{\rm cnt} ]
\cdot {\bf c}^{\rm out} [ {\bf n}_{(M)}^{\rm cnt} ] ,
\\
{\rm Tr} [ \rho_{\rm out}
| \psi_{\rm out}^{{\tilde{\bf d}}{\bf E}} \rangle
\langle \psi_{\rm out}^{{\tilde{\bf d}}{\bf E}}  | ]
&=& \sum_{{\bf n}_{(M)}} P_{\rm PD} [ | {\bf n}_{(M)} \rangle ]
\nonumber \\
& \times &
\left| {\bf c}^{\rm out} [ {\bf n}_{(M)} ]
\cdot {\bf c}^{\rm out} [ {\bf n}_{(M)}^{\rm cnt} ] \right|^2 . \ \ \
\end{eqnarray}
%%%%%
Here, $ {\rm Tr} [ \rho_{\rm out} ] $
is the probability to obtain the expected photon count
$ {\bf n}_{(M)}^{\rm cnt} $
by performing the conditional measurement with this Bell state detector.
The fidelity of manipulation is then evaluated
by considering the sensitivity of photon detector as
%%%%%
\begin{eqnarray}
F [ | \psi_{\rm out}^{{\tilde{\bf d}}{\bf E}} \rangle ]
& \equiv &
1 - \Delta F [ | \psi_{\rm out}^{{\tilde{\bf d}}{\bf E}} \rangle ]
\nonumber \\
&=& 1 - \sum_{(a,b) \not= (0,0)}
f^{(a,b)} [ | \psi_{\rm out}^{{\tilde{\bf d}}{\bf E}} \rangle ]
{\delta \eta}^a \nu^b .
\end{eqnarray}
%%%%%

%%%%%%%%%%
\section{Analysis of efficiencies}
\label{sec:analysis}
%%%%%%%%%%

We can analyze the efficiencies of practical Bell state detectors
and number-state manipulations
by applying the formulas presented so far.

%%%%%
\subsection{Bell state detectors}
%%%%%

In the number-state manipulations based on teleportation,
the number-phase Bell states $ | \phi_- (N,m) \rangle $
in Eq. (\ref{Bell-np1}) may specifically be measured
by the Bell state detectors
\cite{KY-2002,KY-2003,ZPM-2003,ZPM-2004}.
In order to show the efficiency of practical Bell measurement
with linear optics in the present scheme,
we evaluate the confidence typically for the detection
of $ | {\tilde {\bf d}} \rangle = | \phi_- ({\tilde N},0) \rangle $
with number-sum $ {\tilde N} = 1, 2 $ and phase-difference $ m = 0 $.
The number-phase Bell states with nonzero phase-difference $ m $
are also measured similarly by making a phase shift
$ a_2 \rightarrow \omega_{{\tilde N}+1}^m a_2 $
of the mode 2 in Eq. (\ref{Bell-np1})
before the two-mode states enter the Bell state detector.
The Bell state detectors for number-sum $ {\tilde N} = 1, 2 $
are useful for manipulations of qubits and qutrits, as seen later.

The Bell state detector of $ | \phi_- (1,0) \rangle $
with $ {\tilde N} = 1 $
is characterized by the amplitude distribution
and the unitary transformation of optical modes
which are given, respectively, by
%%%%%
\begin{eqnarray}
{\tilde{\bf d}} &=& \frac{1}{\sqrt 2}
\left( \begin{array}{c} 1 \\ 1 \end{array} \right)
\rightarrow | \phi_- ({\tilde N} = 1,0) \rangle ,
\\
U_{\tilde{\bf d}}
&=& \left( \begin{array}{cc}
\frac{1}{\sqrt 2} & - \frac{1}{\sqrt 2} \\
\frac{1}{\sqrt 2} & \frac{1}{\sqrt 2}
\end{array} \right) ,
\label{Ud1}
\end{eqnarray}
%%%%%
where no ancilla is used ($ M = 2 $).
As is well-known, this unitary transformation $ U_{\tilde{\bf d}} $
is realized with a 50:50 beam splitter.
The confidence of this Bell state detector is calculated
in the leading orders of
the expansion with respect to $ \delta \eta $ and $ \nu $ as
%%%%%
\begin{eqnarray}
&{}& \Delta {\cal C}_{\tilde{\bf d}}
\equiv 1 - {\cal C}_{\tilde{\bf d}}
[ | \phi_- ({\tilde N} = 1,0) \rangle ] :
\nonumber \\
&{}& \left[ \begin{array}{ccc}
a & q_{\tilde{\bf d}}^{(a,0)}{\delta \eta}^a
& q_{\tilde{\bf d}}^{(a,1)} \nu {\delta \eta}^a \\
0 & - \! \! \! - \! \! \! - & 1 \\
1 & 3 & - 4 \\
2 & - 3 & 5 \\
3 & 1 & - 2 \\
4 & 0 & 0
\end{array} \right] \ ,
\nonumber
\end{eqnarray}
%%%%%
where the coefficients $ q_{\tilde{\bf d}}^{(a,b)} $ are presented
in this list.
The Bell state detector of $ | \phi_- (2,0) \rangle $
with $ {\tilde N} = 2 $ is characterized by
%%%%%
\begin{eqnarray}
{\tilde{\bf d}} &=& \frac{1}{\sqrt 3}
\left( \begin{array}{c} 1 \\ 1 \\ 1 \end{array} \right)
\rightarrow | \phi_- ({\tilde N} = 2,0) \rangle ,
\\
U_{\tilde{\bf d}}
&=& \left( \begin{array}{ccc}
\frac{1}{\sqrt 2} & 0 & - \frac{1}{\sqrt 2} \\
0 & 1 & 0 \\
\frac{1}{\sqrt 2} & 0 & \frac{1}{\sqrt 2}
\end{array} \right)
\left( \begin{array}{ccc}
1 & 0 & 0 \\
0 & \frac{2}{\sqrt 6} & - \frac{1}{\sqrt 3} \\
0 & \frac{1 + i}{\sqrt 6} & \frac{1 + i}{\sqrt 3}
\end{array} \right)
\nonumber \\
&{}&
\times \left( \begin{array}{ccc}
1 & 0 & 0 \\
0  & \frac{{\sqrt 3}(3 + i)}{4{\sqrt 5}} & - \frac{3 + i}{4}\\
0 & \frac{\sqrt 5}{2{\sqrt 2}}  & \frac{\sqrt 3}{2{\sqrt 2}}
\end{array} \right) ,
\label{Ud2}
\end{eqnarray}
%%%%%
where one ancilla is used ($ M = 3 $)
\cite{KY-2002,KY-2003}.
The confidence of this Bell state detector is calculated
in the leading orders as
%%%%%
\begin{eqnarray}
&{}& \Delta {\cal C}_{\tilde{\bf d}}
\equiv 1 - {\cal C}_{\tilde{\bf d}}
[ | \phi_- ({\tilde N} = 2,0) \rangle ] :
\nonumber \\
&{}& \left[ \begin{array}{ccc}
a & q_{\tilde{\bf d}}^{(a,0)}{\delta \eta}^a
& q_{\tilde{\bf d}}^{(a,1)} \nu {\delta \eta}^a \\
0 & - \! \! \! - \! \! \! - & 7/3 \\
1 & 28/9 & - 304/27 \\
2 & - 1075/324 & 15803/972 \\
3 & 1883/1458 & - 23147/2916 \\
4 & - 2029/26244 & - 19991/39366
\end{array} \right] \ .
\nonumber
\end{eqnarray}
%%%%%

%%%%%
\begin{figure}[t]
\centering
\includegraphics[bb=30 10 510 370,width=.84\linewidth]{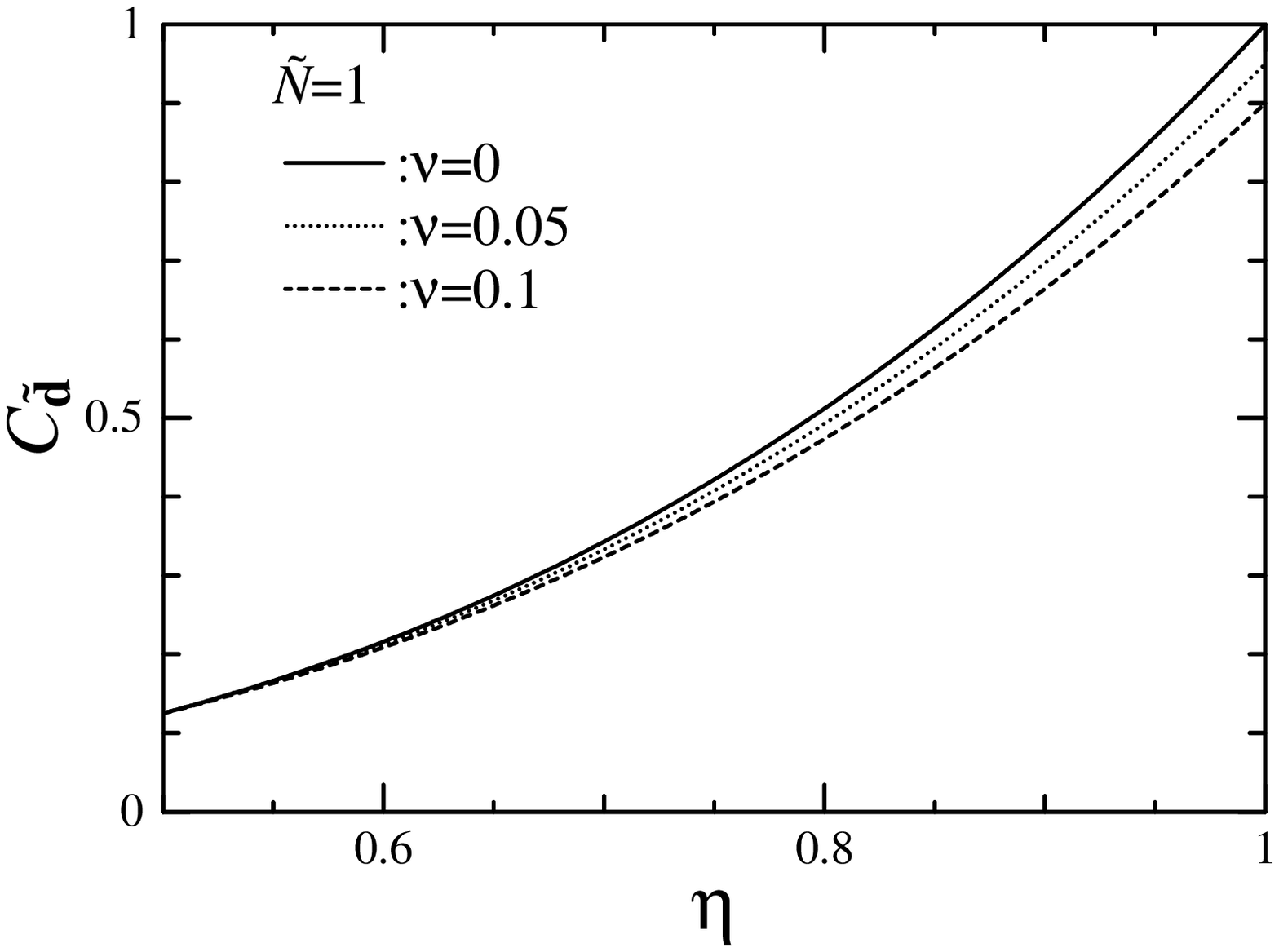}
\caption{The confidence of the detector
to measure the number-phase Bell state
$ | \phi_- ({\tilde N} = 1,0) \rangle $ is shown
depending on $ \eta $ with $ \nu = 0, 0.05, 0.1 $.
}
\vspace*{0.5cm}
\label{confn1}
\end{figure}
%%%%%
%%%%%
\begin{figure}[t]
\centering
\includegraphics[bb=30 10 510 370,width=.84\linewidth]{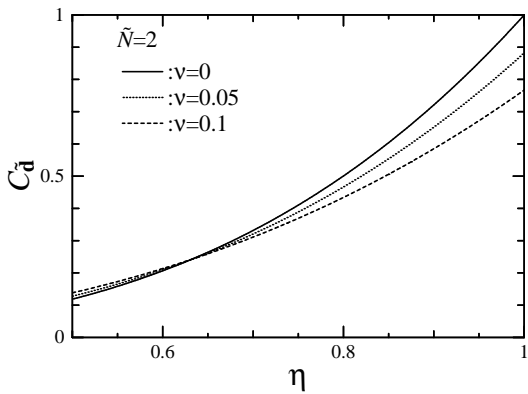}
\caption{The confidence of the detector
to measure the number-phase Bell state
$ | \phi_- ({\tilde N} = 2,0) \rangle $ is shown
depending on $ \eta $ with $ \nu = 0, 0.05, 0.1 $.
}
\label{confn2}
\end{figure}
%%%%%

Numerical estimates of the confidece
are shown in Figs. \ref{confn1} and \ref{confn2}
for $ {\tilde N} = 1 $ and $ {\tilde N} = 2 $, respectively,
depending on $ \eta $ with $ \nu = 0, 0.05, 0.1 $.
Here, it is seen apparently
that the confidences of these Bell state detectors are not so good
unless the quantum efficiency of photon detectors
is rather high as $ \eta > 0.9 $
with the small enough dark count $ \nu $.
It should, however, be remarked that the confidence is defined
in Eq. (\ref{confidence}) with Eq. (\ref{miscount})
to provide a general estimate of Bell state detector,
which is irrespective of the actual contents
of the input two-mode states to be measured.
If the input state contains small components of the states
$ | {\tilde{\bf d}}_\bot \rangle $
other than the desired Bell state $ | {\tilde{\bf d}} \rangle $,
the actual probability to miscount these irrelevant components
as $ | {\tilde{\bf d}} \rangle $ becomes small
according to their portion in the input state.
Furthermore, by the miscount of photon detectors
even the input components of $ | {\tilde{\bf d}}_\bot \rangle $
may contribute to the fidelity to obtain the desired output state.
Hence, the practical Bell measurement may provide high fidelities
for some sorts of number-state manipulations via teleportation,
as seen in the following.

%%%%%
\subsection{Manipulations and preparations}
%%%%%

We next examine some useful manipulations and preparations
of number-states which are based on the teleportation technique
\cite{scissors1,scissors2,scissors3,KY-2002,KY-2003,ZPM-2003,ZPM-2004};
scissors, reversal, generalized number-phase Bell state
and truncated maximally squeezed vacuum state.
This analysis of efficiencies will indeed be relevant
for feasible experimental realizations
of these sorts of operations particularly for qubits and qutrits.
The success probabilities have been calculated
by assuming the ideal Bell state detectors in Ref. \cite{KY-2003},
which provide approximate estimates even in the present scheme
utilizing realistic photon detectors with reasonable efficiency.
The precise evaluations of success probabilities
can be made by applying the formulas presented
in Secs. \ref{sec:bell-detector} and \ref{sec:manipulation}
for the practical Bell state detectors.
A detailed analysis may be reserved for a future study,
while it is not the aim of the present work.

The teleportation based manipulations are specified
by the sets of input state, EPR resource and Bell measurement as
%%%%%
\begin{equation}
{\cal S} [ {\rm manipulation} ]
= \{ | \psi_{\rm in} \rangle ,
| {\rm EPR} \rangle , | {\tilde{\bf d}} \rangle \} .
\end{equation}
%%%%%
Specifically, we take the number-phase Bell measurement
of $ | {\tilde{\bf d}} \rangle
= | \phi_- (N,0) \rangle $ ($ r=1 $)
with $ {\tilde N} = N = 1, 2 $ and $ m = 0 $.
As for the EPR resources,
we take the two-mode squeezed vacuum state $ | \lambda \rangle $
with squeezing parameter $ \lambda < 1 $,
the generalized number-phase Bell state $ | \phi_- (N,0,r) \rangle $
and the truncated maximally squeezed vacuum state
$ | \lambda = 1 , N \rangle $, which is given by
%%%%%
\begin{equation}
| \lambda = 1 , N \rangle
= \frac{| 0 \rangle | 0 \rangle + | 1 \rangle | 1 \rangle
+ \ldots + | N \rangle | N \rangle}{\sqrt{N+1}} .
\end{equation}
%%%%%
The squeezed vacuum state is taken as the primary resource of entanglement,
and the other EPR resources can be prepared in the present scheme as
%%%%%
\begin{eqnarray}
| \lambda \rangle \rightarrow | \phi_- (N,0,r) \rangle
\rightarrow | \lambda = 1 , N \rangle ,
\nonumber
\end{eqnarray}
%%%%%
which will be described below.

The ingredients for the relevant manipulations and preparations
are listed as follows.
%%%%%
\begin{flushleft}
$ \bullet $ {\bf Scissors}:
\end{flushleft}
%%%%%
\begin{eqnarray}
&& | \psi_{\rm in}^{(1)} \rangle
= \sum_{n=0}^\infty c^{\rm in}_n | n \rangle
\rightarrow \sum_{n=0}^N c^{\rm in}_n | n \rangle ,
\nonumber \\
{} \nonumber \\
&& {\cal S} [ {\rm scissors} ]
= \{ | \psi_{\rm in}^{(1)} \rangle ,
| \phi_- (N,0) \rangle ,
| \phi_- (N,0) \rangle \} .
\nonumber
\end{eqnarray}
%%%%%

%%%%%
\begin{flushleft}
$ \bullet $ {\bf Reversal}:
\end{flushleft}
%%%%%
\begin{eqnarray}
&& | \psi_{\rm in}^{(1)} \rangle
= \sum_{n=0}^\infty c^{\rm in}_n | n \rangle
\rightarrow \sum_{n=0}^N c^{\rm in}_{N - n} | n \rangle ,
\nonumber \\
{} \nonumber \\
&& {\cal S} [ {\rm reversal} ]
= \{ | \psi_{\rm in}^{(1)} \rangle ,
| \lambda = 1 , N \rangle , | \phi_- (N,0) \rangle \} .
\nonumber
\end{eqnarray}
%%%%%

%%%%%
\begin{flushleft}
$ \bullet $ {\bf Generalized number-phase Bell state}:
\end{flushleft}
%%%%%
\begin{eqnarray}
&& | \lambda \rangle \rightarrow | \phi_- (N,0,r) \rangle ,
\nonumber \\
{} \nonumber \\
&& {\cal S} [ | \phi_- (N,0,r) \rangle ]
= \{ | \lambda \rangle ,
| \lambda^\prime = r \lambda \rangle ,
| \phi_- (N,0) \rangle \} .
\nonumber
\end{eqnarray}
%%%%%

%%%%%
\begin{flushleft}
$ \bullet $ {\bf Truncated maximally squeezed vacuum state}:
\end{flushleft}
%%%%%
\begin{eqnarray}
&& | \phi_- (N,0, 1 / \lambda ) \rangle
\rightarrow | \lambda = 1 , N \rangle ,
\nonumber \\
{} \nonumber \\
&& {\cal S} [ | \lambda = 1 , N \rangle ]
= \{ | \phi_- (N,0,1 / \lambda \rangle ,
| \lambda \rangle , | \phi_- (N,0) \rangle \} .
\nonumber
\end{eqnarray}
%%%%%
The matrices representing the relevant EPR resources are given as follows.
The squeezed vacuum state is represented by
%%%%%
\begin{eqnarray}
{\bf E} [ | \lambda \rangle ]
= {\sqrt{1 - \lambda^2}}
\left( \begin{array}{ccccc}
1 & 0 & 0 & 0 & \cdots \\
0 & \lambda & 0 & 0 & \cdots \\
0 & 0 & \lambda^2 & 0 & \cdots \\
0 & 0 & 0 & \lambda^3 & \cdots \\
\vdots & \vdots & \vdots & \vdots & \ddots
\end{array} \right) .
\nonumber
\end{eqnarray}
%%%%%
The truncated maximally squeezed vacuum states are represented
for $ N = 1 $ and $ 2 $, respectively, by
%%%%%
\begin{eqnarray}
{\bf E} [ | \lambda = 1 , N = 1 \rangle ]
&=& \left( \begin{array}{ccccc}
\frac{1}{\sqrt 2} & 0 & 0 & 0 & \cdots \\
0 & \frac{1}{\sqrt 2} & 0 & 0 & \cdots \\
0 & 0 & 0 & 0 & \cdots \\
0 & 0 & 0 & 0 & \cdots \\
\vdots & \vdots & \vdots & \vdots & \ddots
\end{array} \right) ,
\nonumber \\
{\bf E} [ | \lambda = 1 , N = 2 \rangle ]
&=& \left( \begin{array}{ccccc}
\frac{1}{\sqrt 3} & 0 & 0 & 0 & \cdots \\
0 & \frac{1}{\sqrt 3} & 0 & 0 & \cdots \\
0 & 0 & \frac{1}{\sqrt 3} & 0 & \cdots \\
0 & 0 & 0 & 0 & \cdots \\
\vdots & \vdots & \vdots & \vdots & \ddots
\end{array} \right) .
\nonumber
\end{eqnarray}
%%%%%
The generalized number-phase Bell states ($ m = 0 $) are represented
for $ N = 1 $ and $ 2 $, respectively, by
%%%%%
\begin{eqnarray}
{\bf E} [ | \phi_- (N = 1,0,r) \rangle ]
&=& D(1,r)
\left( \begin{array}{ccccc}
0 & r & 0 & 0 & \cdots \\
1 & 0 & 0 & 0 & \cdots \\
0 & 0 & 0 & 0 & \cdots \\
0 & 0 & 0 & 0 & \cdots \\
\vdots & \vdots & \vdots & \vdots & \ddots
\end{array} \right) ,
\nonumber \\
{\bf E} [ | \phi_- (N = 2,0,r) \rangle ]
&=& D(2,r)
\left( \begin{array}{ccccc}
0 & 0 & r^2 & 0 & \cdots \\
0 & r & 0 & 0 & \cdots \\
1 & 0 & 0 & 0 & \cdots \\
0 & 0 & 0 & 0 & \cdots \\
\vdots & \vdots & \vdots & \vdots & \ddots
\end{array} \right) .
\nonumber
\end{eqnarray}
%%%%%

%%%%%
\subsubsection{Scissors and reversal}
%%%%%

In order to show the efficiency of number-state manipulations
with the practical Bell state detectors,
we evaluate the fidelity of the scissors and reversal
for qubit ($ N = 1 $) and qutrit ($ N = 2 $).
A coherent state may be taken typically as the input,
%%%%%
\begin{eqnarray}
| \alpha \rangle = {\rm e}^{- | \alpha |^2 / 2}
\sum_{n=0}^\infty \frac{\alpha}{\sqrt{n!}} | n \rangle
\nonumber
\end{eqnarray}
%%%%%
with
%%%%%
\begin{eqnarray}
{\bf c}^{\rm in}
= {\rm e}^{- | \alpha |^2 / 2}
( 1 , \alpha , \alpha^2 / {\sqrt{2}} , \ldots )^{\rm T} .
\nonumber
\end{eqnarray}
%%%%%
Then, by using particularly the $ N = 2 $ scissors
we can prepare a qutrit
%%%%%
\begin{eqnarray}
| \psi ({\mbox{qutrit1}}) \rangle
= | 0 \rangle + \alpha | 1 \rangle + ( \alpha^2 / {\sqrt{2}} ) | 2 \rangle ,
\nonumber
\end{eqnarray}
%%%%%
while we can rearrange this qutrit by the reversal as
%%%%%
\begin{eqnarray}
| \psi ({\mbox{qutrit2}}) \rangle
= ( \alpha^2 / {\sqrt{2}} ) | 0 \rangle + \alpha | 1 \rangle + | 2 \rangle ,
\nonumber
\end{eqnarray}
%%%%%
where the normalization factors are omitted.

By applying the formulas presented in Secs. \ref{sec:bell-detector}
and \ref{sec:manipulation},
the fidelity of the scissors is calculated straightforwardly,
which is give in the leading orders
for $ N = 1 $ and $ 2 $, respectively,
with $ | \alpha | = {\sqrt 3} $ for example as
%%%%%
\begin{eqnarray}
&{}& \Delta F_{\rm SC}
\equiv 1 - F_{\rm SC} [ N = 1 ; | \alpha | = {\sqrt 3} ] :
\nonumber \\
&{}& \left[ \begin{array}{ccc}
a & f^{(a,0)}{\delta \eta}^a
& f^{(a,1)} \nu {\delta \eta}^a \\
0 & - \! \! \! - \! \! \! - & 1/8 \\
1 & 9/16 & 1/8 \\
2 & - 27/64 & 7/128 \\
3 & 81/256 & 41/256 \\
4 & - 243/1024 & 85/2048
\end{array} \right] \ ,
\nonumber
\end{eqnarray}
%%%%%
%%%%%
\begin{eqnarray}
&{}& \Delta F_{\rm SC}
\equiv 1 - F_{\rm SC} [ N = 2 ; | \alpha | = {\sqrt 3} ] :
\nonumber \\
&{}& \left[ \begin{array}{ccc}
a & f^{(a,0)}{\delta \eta}^a
& f^{(a,1)} \nu {\delta \eta}^a \\
0 & - \! \! \! - \! \! \! - & 7/17 \\
1 & 483/1156 & - 4826/4913 \\
2 & 1431/4624 & 48021/78608 \\
3 & - 235683/314432 & 1276203/2672672 \\
4 & 1443321/2515456 & - 36559049/21381376
\end{array} \right] \ .
\nonumber
\end{eqnarray}
%%%%%
%%%%%
\begin{figure}[t]
\centering
\includegraphics[bb=30 10 510 370,width=.84\linewidth]{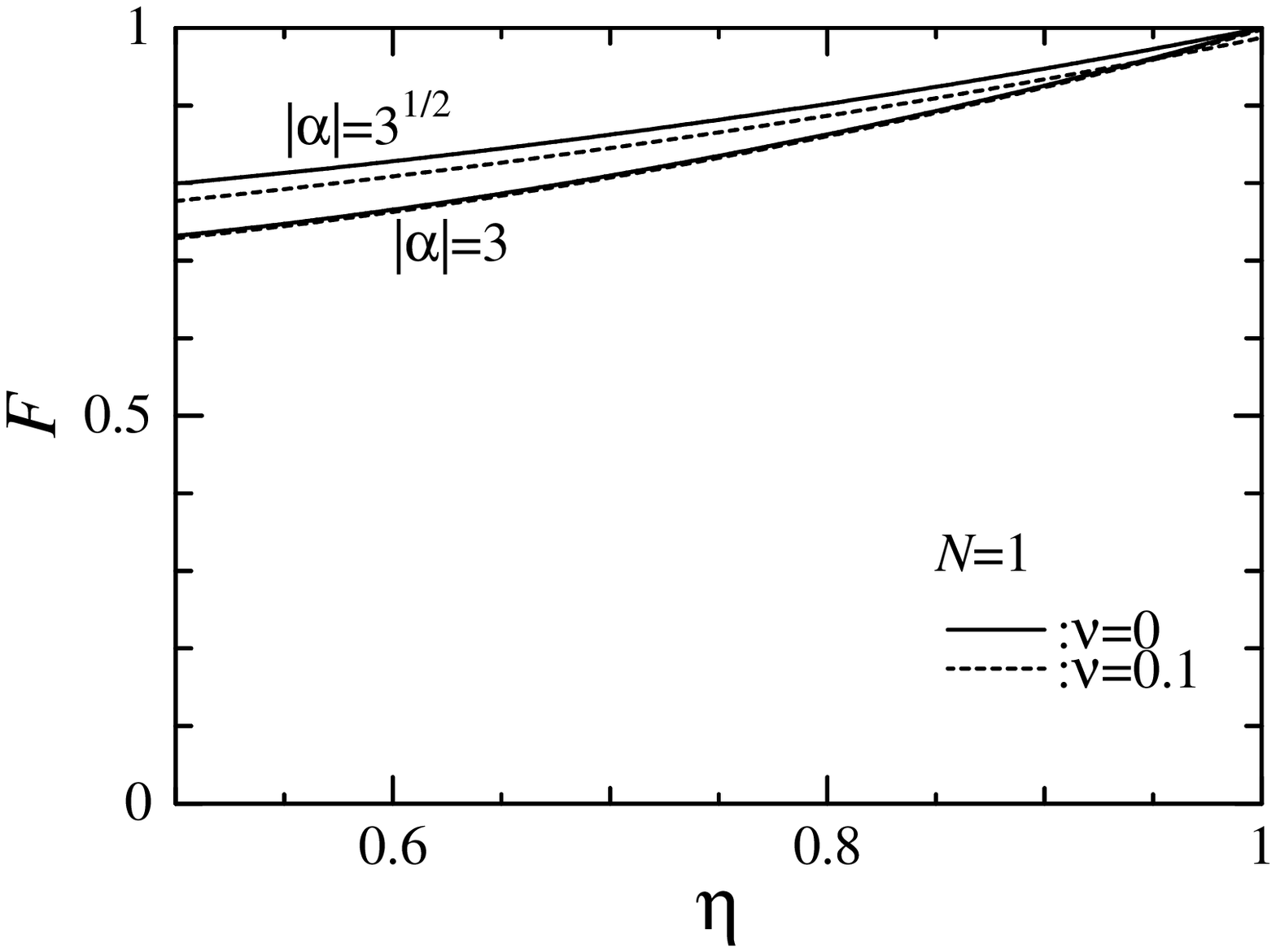}
\caption{The fidelity of the scissors and reversal
$ F_{\rm SC} = F_{\rm RV} $ with $ N = 1 $
for the input coherent state with $ | \alpha | = {\sqrt 3} $
and $ | \alpha | = 3 $ is shown depending on $ \eta $ with $ \nu = 0, 0.1 $.
}
\label{revn1}
\vspace*{0.5cm}
\end{figure}
%%%%%
%%%%%
\begin{figure}[t]
\centering
\includegraphics[bb=30 10 510 370,width=.84\linewidth]{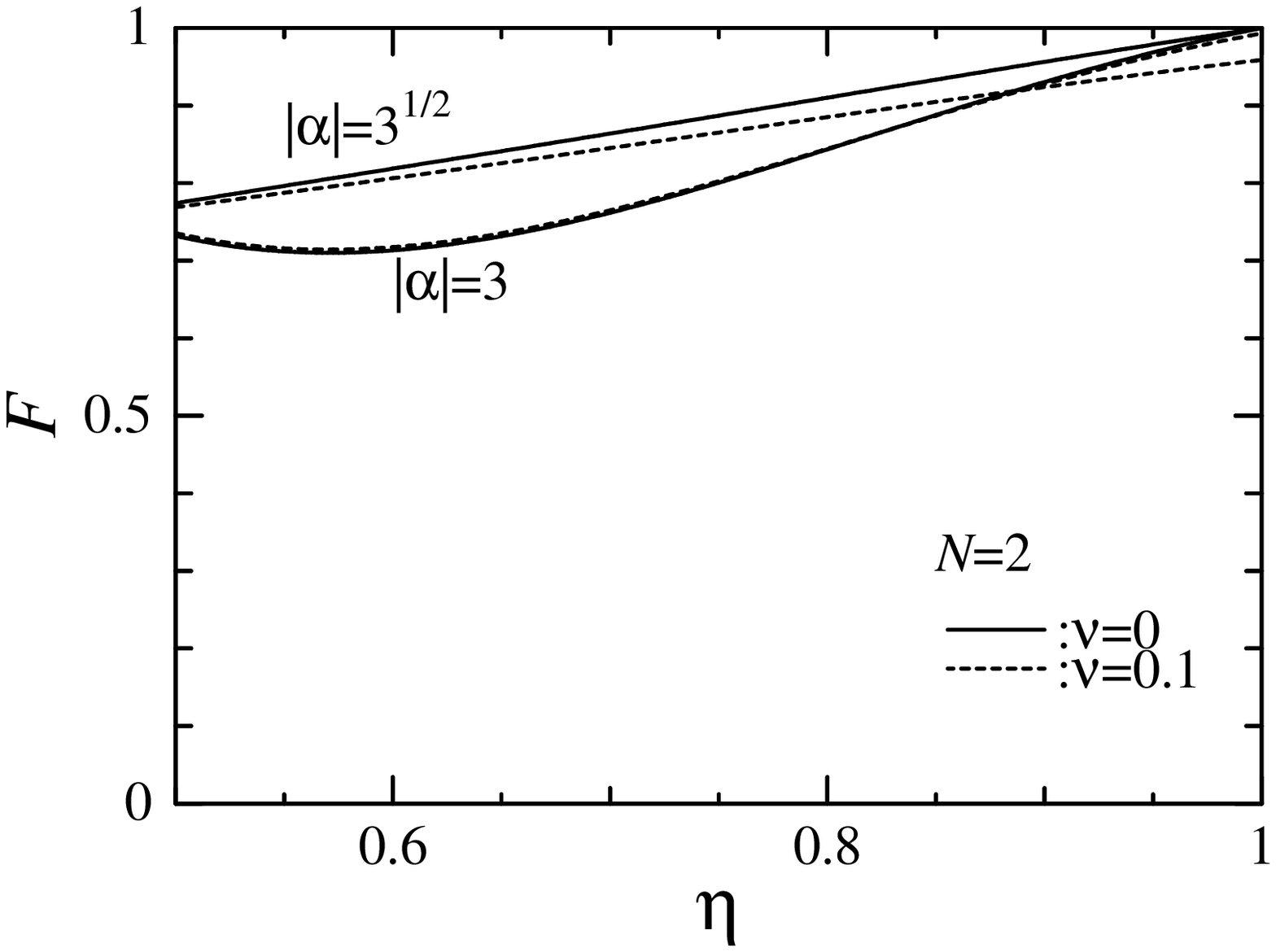}
\caption{The fidelity of the scissors and reversal
$ F_{\rm SC} = F_{\rm RV} $ with $ N = 2 $
for the input coherent state with $ | \alpha | = {\sqrt 3} $
and $ | \alpha | = 3 $ is shown depending on $ \eta $ and $ \nu = 0, 0.1 $.
}
\label{revn2}
\end{figure}
%%%%%
The fidelity of the reversal $ F_{\rm RV} $ is the same
as that of the scissors $ F_{\rm SC} $ in the present scheme
(if the EPR resources are ideally prepared):
%%%%%
\begin{equation}
F_{\rm RV} = F_{\rm SC} .
\end{equation}
%%%%%
This is verified by the relation
%%%%%
\begin{eqnarray}
{\bf E}^\dagger {\bf E} \ [{\rm SC}]
= {\bf E}^\dagger {\bf E} \ [{\rm RV}]
\nonumber
\end{eqnarray}
%%%%%
for $ | \phi_- (N,0) \rangle $ [SC]
and $ | \lambda = 1 , N \rangle $ [RV]
in calculating the inner product of the output states
with Eqs. (\ref{c-out}) and (\ref{TdE}),
%%%%%
\begin{eqnarray}
&{}& \langle \psi_{\rm out} [ {\bf n}^\prime_{(M)} ] |
| \psi_{\rm out} [ {\bf n}_{(M)} ] \rangle
= {\bf c}^{\rm out} [ {\bf n}^\prime_{(M)} ] \cdot
{\bf c}^{\rm out} [ {\bf n}_{(M)} ]
\nonumber \\
&{}& \hspace{0.5cm} = {\bf c}^{{\rm in} \dagger}
{\bf B}^{{\tilde{\bf d}} \dagger} [ {\bf n}^\prime_{(M)} ]
{\bf R}_{N^\prime}^\dagger {\bf E}^\dagger
{\bf E} {\bf R}_N {\bf B}^{\tilde{\bf d}} [ {\bf n}_{(M)} ] 
{\bf c}^{\rm in} .
\nonumber
\end{eqnarray}
%%%%%

Numerical estimates of the fidelity of the scissors and reversal
are shown in Figs. \ref{revn1} and \ref{revn2}
for $ N = 1 $ and $ N = 2 $, respectively,
depending on $ \eta $ with $ \nu = 0, 0.1 $.
It is here noticed in Fig. \ref{revn2}
that the fidelity is apparently increasing
for $ \eta \lesssim 0.6 $
in the case of $ N = 2 $ and $ | \alpha | = 3 $.
This would indicate that the approximation
with the leading terms up to $ {\delta \eta}^4 $
is not good enough with $ {\delta \eta}^5 \sim 0.01 $
for $ {\delta \eta} \sim 0.4 $
in the case of considerably large $ | \alpha | $.

It is already shown that a high fidelity is achievable
in the $ N = 1 $ scissors for a small enough amplitude
of the input coherent state,
e.g., $ F_{\rm SC} > 0.9 $ for $ \eta = 0.7 $ with $ | \alpha | = 1 $,
where the conventional photon detectors
resolving one or more photons may be used
\cite{scissors3}.
In the present scheme including the cases of $ N \geq 2 $,
we should use the single-photon detectors,
which resolve zero, one or more photons,
since two or more photons may enter some of the detectors.
The best available single-photon detector
provides the quantum efficiency $ \eta \approx 0.7 - 0.9 $
\cite{SPD1}.
Its dark count rate is roughly given
as $ R_{\rm dark} \sim 10^4 {\rm s}^{-1} $.
Then, the mean dark count is estimated
as $ \nu = \tau_{\rm res} R_{\rm dark} \sim 10^{-4} $
by assuming the detector resolution time $ \tau_{\rm res} = 10 {\rm ns} $
\cite{scissors3}.
It is encouraging for future experimental attempts
that new proposals have been made for single-photon detection
to achieve the quantum efficiency close to unity
\cite{SPD2}.

In Figs. \ref{revn1} and \ref{revn2},
we present the estimates of fidelity by taking somewhat large amplitudes
as $ | \alpha | = {\sqrt 3} $ and $ | \alpha | = 3 $
to emphasize the effect of imperfectness of single-photon detectors.
A high fidelity can really be obtained for example as
$ F_{\rm SC} = F_{\rm RV} \gtrsim 0.9 $
for $ | \alpha | = {\sqrt 3} $ with $ \eta \geq 0.8 $
in the scissors and reversal of $ N $ = 1 and 2.
If a smaller amplitude is taken as $ | \alpha | < 1 $,
the fidelity becomes higher, as seen in Ref. \cite{scissors3}.
It should be noted here that the actual fidelities of scissors and reversal
are slightly decreased by those for preparing
the EPR resources $ | \phi_- (N,0) \rangle $
and $ | \lambda = 1 , N \rangle $,
which can be higher than 0.95 for $ \eta \geq 0.7 $
with the small enough squeezing parameters $ \sim 0.1 $, as estimated later.

A spuriouly large $ \nu = 0.1 $ is taken
in Figs. \ref{revn1} and \ref{revn2}
so as to make the correction by the dark count visible.
Actually, the effect of the dark count is fairly small
in these scissors and reversal of $ N = 1, 2 $,
since $ | f^{(a,1)} | \nu \sim 10^{-4} $
with $ | f^{(a,1)} | \lesssim 1 $ for the reasonable $ \nu \sim 10^{-4} $.
It should, however, be remarked that the fidelity
for preparing the EPR resource $ | \phi_- (N,0) \rangle $
is somewhat sensitive to the dark count $ \nu $
providing a correction $ \sim 0.005 $, as seen later.

The net success probabilities for the scissors and reversal
are roughly given
from the estimates in the case of ideal Bell state detectors
\cite{KY-2003} as
%%%%%
\begin{eqnarray}
P_{\rm SC} (N) \sim \frac{p(N)^2}{(N+1)^2} \lambda^{2N} ,
P_{\rm RV} (N) \sim \frac{p(N)^3}{(N+1)^4} \lambda^{\prime 2N} ,
\nonumber
\end{eqnarray}
%%%%%
where $ \lambda^2 , \lambda^{\prime 2} \ll 1 $
for the squeezing parameters relevant for preparing the EPR resources.
Henceforth $ p(N) $ represents the success probability
of the ideal measurement of $ | \phi_- (N,0) \rangle $, e.g.,
$ p(1) = 1 $ and $ p(2) = 1/2 $ for the Bell state detectors
presented so far.
(Note that $ p(2) = 3/8 $ was given in error in Ref. \cite{KY-2003}.)
The success probabilities to prepare
the EPR resources $ | \phi_- (N,0) \rangle $
and $ | \lambda = 1 , N \rangle $ are included
in the above estimates for the scissors and reversal, respectively.
It appears that $ P_{\rm RV} (N) $ is rather suppressed,
since an additional Bell measurement is made to prepare
$ | \lambda = 1 , N \rangle $ from $ | \lambda \rangle $
and $ | \phi_- ( N , 0 , 1 / \lambda ) \rangle $.
That is, $ | \lambda = 1 , N \rangle $ is generated
from three squeezed vacuum states by making the Bell measurement twice.
Numerically, by taking typically $ \lambda = \lambda^\prime = 1/4 $
we have $ P_{\rm SC} (1) \sim 2 \times 10^{-2} $,
$ P_{\rm SC} (2) \sim 1 \times 10^{-4} $
and $ P_{\rm RV} (1) \sim 4 \times 10^{-3} $,
$ P_{\rm RV} (2) \sim 6 \times 10^{-6} $.

%%%%%
\subsubsection{Generalized number-phase Bell states}
%%%%%

%%%%%
\begin{figure}[t]
\centering 
\includegraphics[bb=30 10 510 370,width=.84\linewidth]{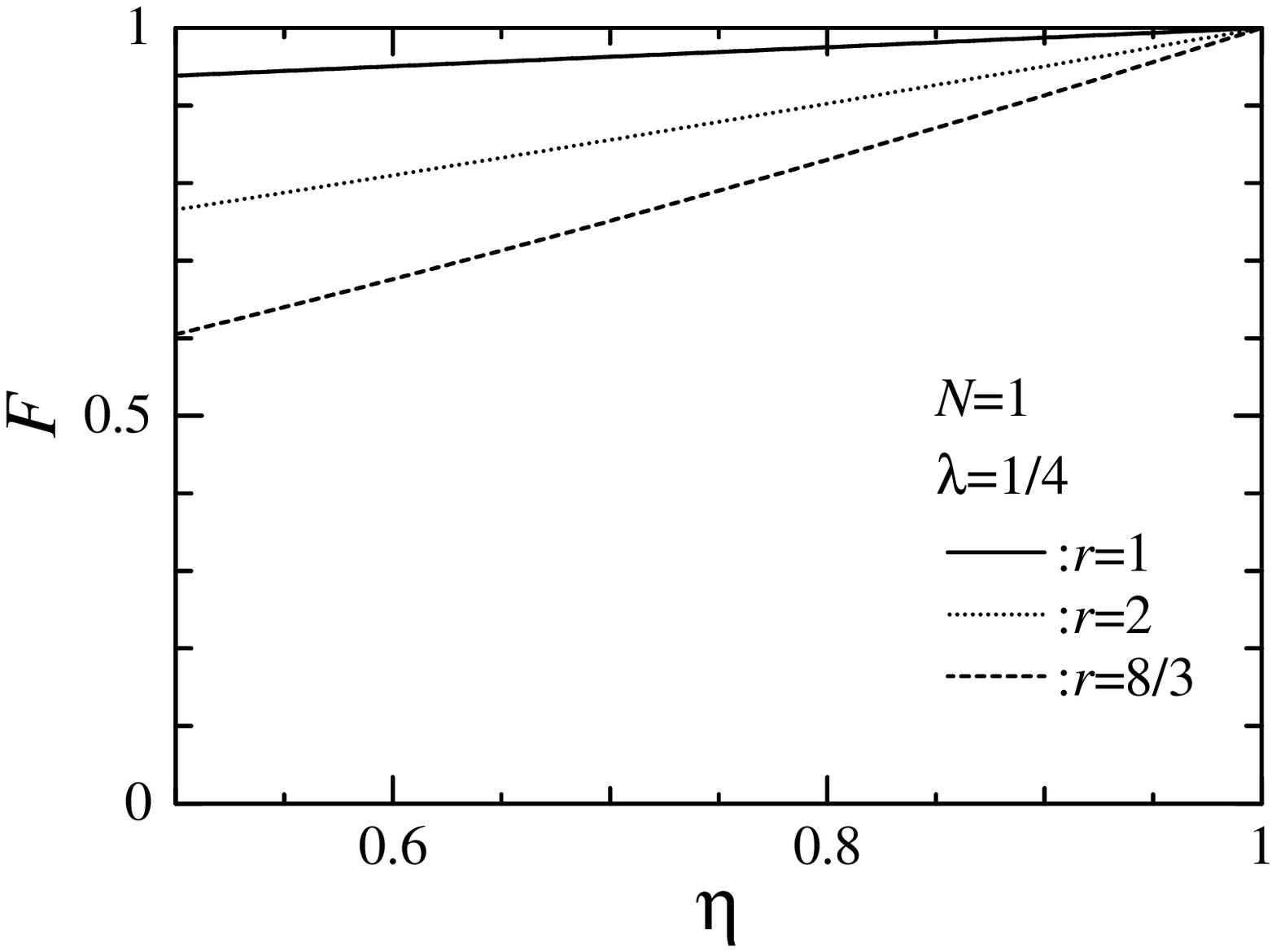}
\caption{The fidelity $ F_{\rm GB} $ for the preparation
of $ N = 1 $ generalized number-phase Bell state
$ | \phi_- (N = 1,0,r) \rangle $
is shown depending on $ \eta $ with $ \nu = 0 $.
Here, $ \lambda = 1/4 $ is taken for the input state,
and then $ \lambda^\prime = r \lambda $ of the EPR resource
is given with some typical values of $ r $.
}
\vspace*{0.5cm}
\label{gbn1}
\end{figure}
%%%%%

%%%%%
\begin{figure}[t]
\centering
\includegraphics[bb=30 10 510 370,width=.84\linewidth]{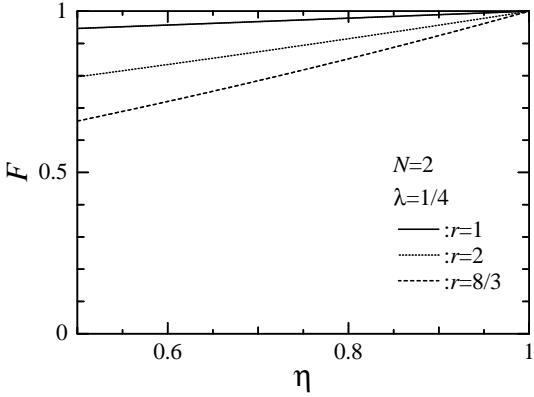}
\caption{The fidelity $ F_{\rm GB} $ for the preparation
of $ N = 2 $ generalized number-phase Bell state
$ | \phi_- (N = 2,0,r) \rangle $
is shown depending on $ \eta $ with $ \nu = 0 $.
Here, $ \lambda = 1/4 $ is taken for the input state,
and then $ \lambda^\prime = r \lambda $ of the EPR resource
is given with some typical values of $ r $.
}
\label{gbn2}
\end{figure}
%%%%%

We next consider the preparation of two-mode entangled states.
For the preparation of the generalized number-phase Bell state
$ | \phi_- (N,0,r) \rangle $,
a squeezed vacuum $ | \lambda^\prime = r \lambda \rangle $
is used as the EPR resource,
and another squeezed vacuum $ | \lambda \rangle $
is taken as the input state, which is represented by the matrix
%%%%%
\begin{eqnarray}
{\bf c}^{\rm in} = {\bf E} [ | \lambda \rangle ] .
\nonumber
\end{eqnarray}
%%%%%
Then, by applying the formulas in Secs. \ref{sec:bell-detector}
and \ref{sec:manipulation},
the fidelity is calculated for example
with $ \lambda = \lambda^\prime = 1/4 $ ($ r = 1 $)
for the input state and EPR resource as
%%%%%
\begin{eqnarray}
&{}& \Delta F_{\rm GB} \equiv 1 - F_{\rm GB}
[ N = 1 ; r = 1 , \lambda = \lambda^\prime = 1/4 ] :
\nonumber \\
&{}& \left[ \begin{array}{ccc}
a & f^{(a,0)}{\delta \eta}^a
& f^{(a,1)} \nu {\delta \eta}^a \\
0 & - \! \! \! - \! \! \! - & 32 \\
1 & 1/8 & 24 \\
2 & - 1/256 & 197/8 \\
3 & 0 & 1575/64 \\
4 & 0 & 1575/64
\end{array} \right] \ ,
\nonumber
\end{eqnarray}
%%%%%
%%%%%
\begin{eqnarray}
&{}& \Delta F_{\rm GB} \equiv 1 - F_{\rm GB}
[ N = 2 ; r = 1 , \lambda = \lambda^\prime = 1/4 ] :
\nonumber \\
&{}& \left[ \begin{array}{ccc}
a & f^{(a,0)}{\delta \eta}^a
& f^{(a,1)} \nu {\delta \eta}^a \\
0 & - \! \! \! - \! \! \! - & 56 \\
1 & 7/64 & 175/4 \\
2 & - 49/12286 & 17143/384 \\
3 & 343/7077888 & 19736731/442368 \\
4 & 0 & 842106125/18874368
\end{array} \right] \ .
\nonumber
\end{eqnarray}
%%%%%

Numerical estimates of the fidelity
are shown in Figs. \ref{gbn1} and \ref{gbn2}
for $ N = 1 $ and $ N = 2 $, respectively,
depending on $ \eta $ with $ \nu = 0 $ for simplicity.
Here, $ \lambda = 1/4 $ is taken for the input state,
and then $ \lambda^\prime = r \lambda $ of the EPR resource
is given with some typical values of $ r $.
A higher fidelity is obtained for a smaller $ r < 1 $,
though it is not depicted in these figures.
It is in fact checked that the coefficients
$ f^{(a,0)} $ and $ f^{(a,1)} $ for $ F_{\rm GB} $
are calculated to be independent of the squeezing parameter
$ \lambda $ of the input state.
That is, they are determined solely
by the squeezing parameter $ \lambda^\prime $ of the EPR resource.
This may be ascribed to the fact
that the optical setup of the present Bell state detector
with $ U_{\tilde{\bf d}} $ in Eqs. (\ref{Ud1}) and (\ref{Ud2})
and $ {\bf n}_{(M)}^{\rm cnt} $ in Eq. (\ref{photon-count})
is asymmetric under the exchange of the input modes 1 and 2, i.e.,
in this case $ | \lambda \rangle \leftrightarrow | \lambda^\prime \rangle $.
Then, by taking the small enough $ \lambda^\prime \leq 1/4 $
a fairly high fidelity $ F_{\rm GB} > 0.95 $
can be achieved for $ \eta \geq 0.7 $ and $ \nu \sim 10^{-4} $.
As for the effect of the dark count, the fidelity $ F_{\rm GB} $
for preparing $ | \phi_- (N,0,r) \rangle $
appears somewhat sensitive to $ \nu $.
It provides a correction estimated as $ | f^{(a,1)} | \nu \sim 0.005 $
with $ | f^{(a,1)} | \lesssim 50 $ for the reasonable $ \nu \sim 10^{-4} $.

The success probability to prepare $ | \phi_- (N,0,r) \rangle $
is estimated roughly \cite{KY-2003} as
%%%%%
\begin{eqnarray}
P_{\rm GB} (N) \sim \left\{ \begin{array}{ll}
p(N) {\bar \lambda}^{2N} / (N+1) & ( r^2 \gg 1 , r^2 \ll 1 ) \\
p(N) \lambda^{2N} & ( r \approx 1 )
\end{array} \right. ,
\nonumber
\end{eqnarray}
%%%%%
where $ {\bar \lambda} = {\rm max}[ \lambda , \lambda^\prime ] $.
Numerically, for example we have
$ P_{\rm GB} (1) \sim 3 \times 10^{-2} $
and $ P_{\rm GB} (2) \sim 1 \times 10^{-3} $
with $ \lambda^\prime = 1/4 > \lambda $.

%%%%%
\subsubsection{Truncated maximally squeezed vacuum states}
%%%%%

%%%%%
\begin{figure}[t]
\centering 
\includegraphics[bb=30 10 510 370,width=.84\linewidth]{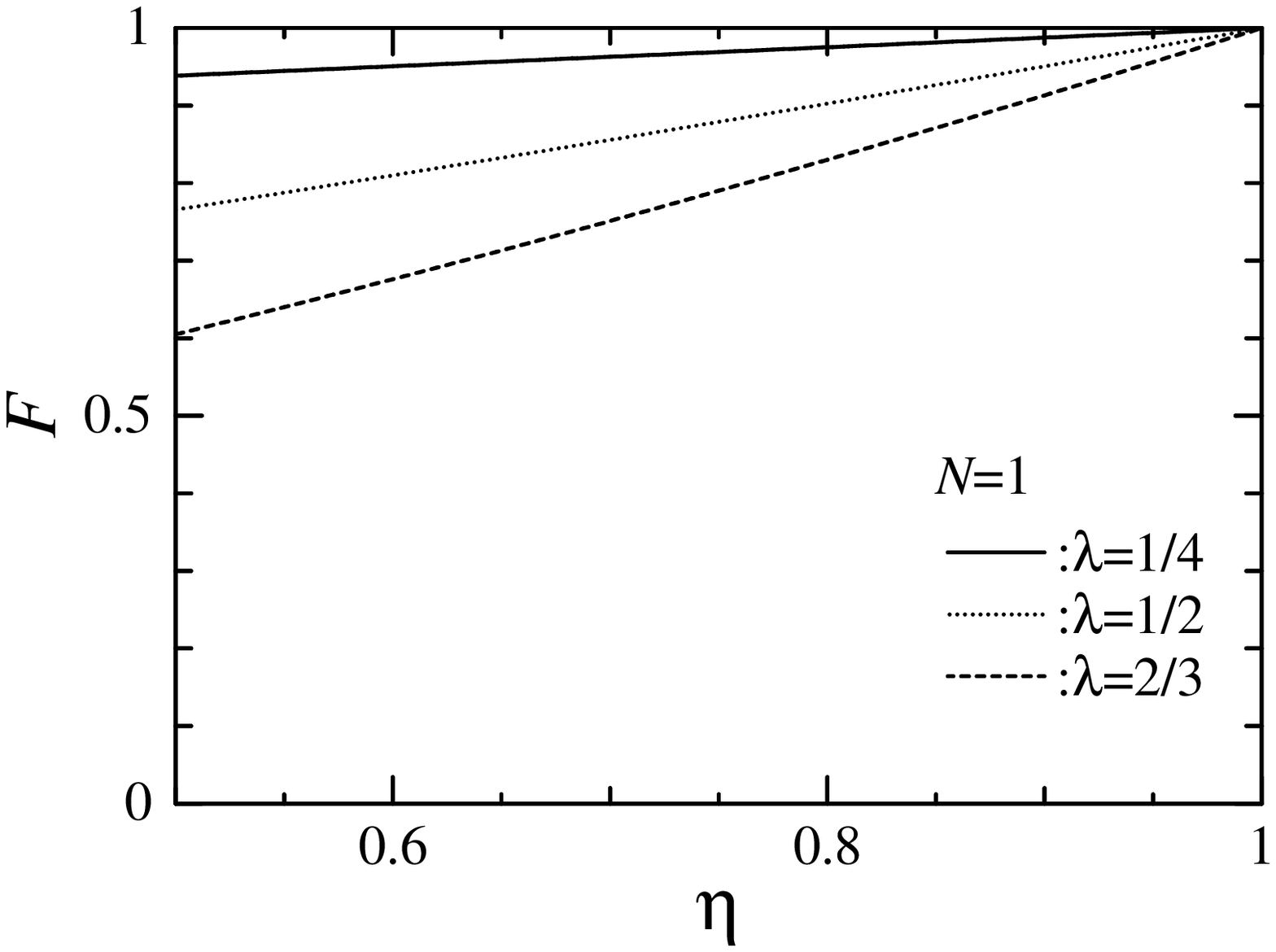}
\caption{The fidelity $ F_{\rm MSV} $ for the preparation
of $ N = 1 $ truncated maximally squeezed vacuum state
$ | \lambda = 1 , N = 1 \rangle $
is shown depending on $ \eta $ with $ \nu = 0 $.
Some typical values are taken for the relevant squeezing parameter
$ \lambda $.
}
\vspace*{0.5cm}
\label{msvn1}
\end{figure}
%%%%%
%%%%%
\begin{figure}[t]
\centering
\includegraphics[bb=30 10 510 370,width=.84\linewidth]{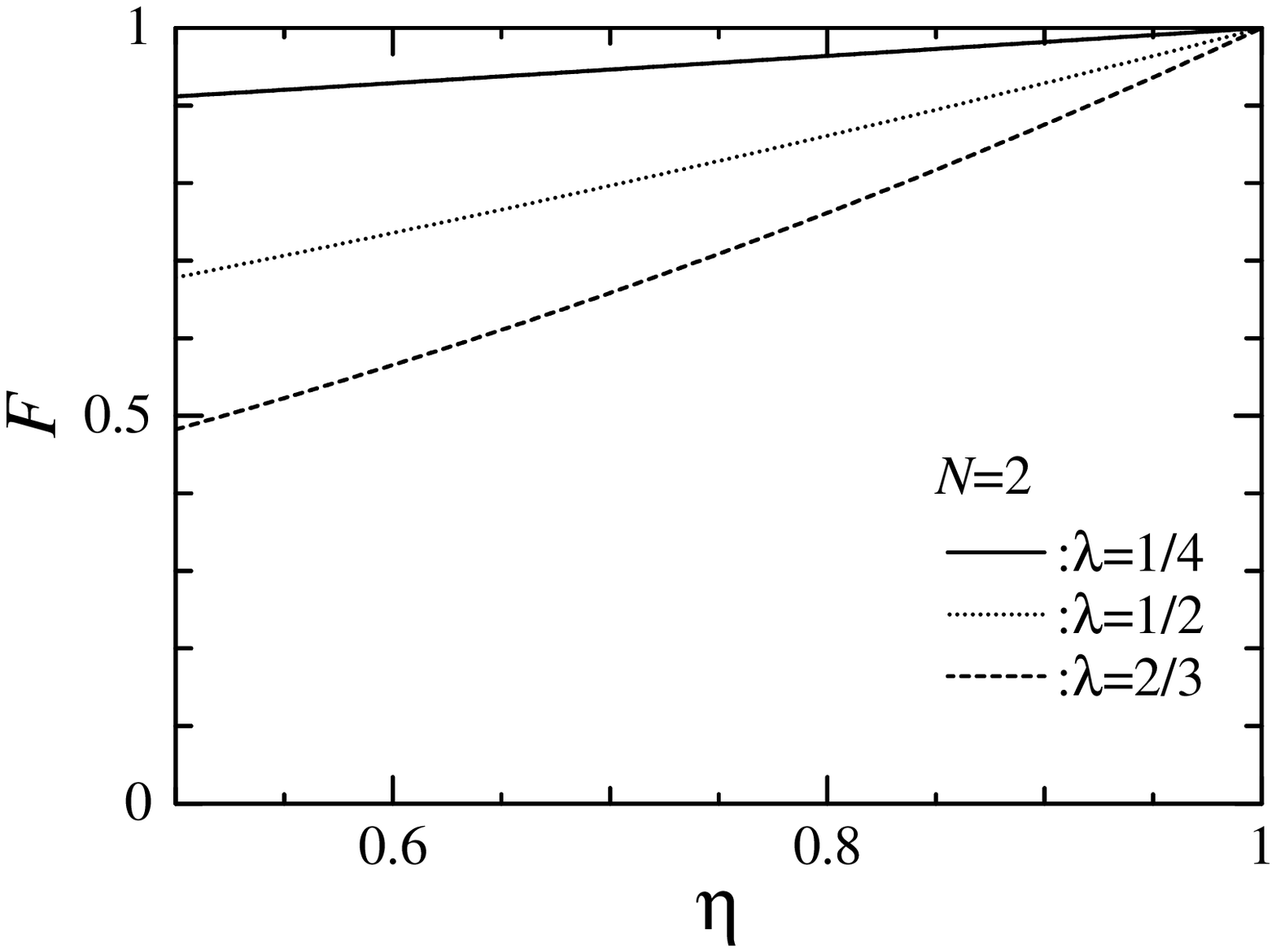}
\caption{The fidelity $ F_{\rm MSV} $ for the preparation
of $ N = 2 $ truncated maximally squeezed vacuum state
$ | \lambda = 1 , N = 2 \rangle $
is shown depending on $ \eta $ with $ \nu = 0 $.
Some typical values are taken for the relevant squeezing parameter
$ \lambda $.
}
\label{msvn2}
\end{figure}
%%%%%

For generating the truncated maximally squeezed vacuum states,
%%%%%
\begin{eqnarray}
&{}& | \lambda = 1 , N = 1 \rangle
= ( | 0 \rangle | 0 \rangle + | 1 \rangle | 1 \rangle )/{\sqrt 2} ,
\nonumber \\
&{}& | \lambda = 1 , N = 2 \rangle
= ( | 0 \rangle | 0 \rangle + | 1 \rangle | 1 \rangle
+ | 2 \rangle | 2 \rangle )/{\sqrt 3} ,
\nonumber
\end{eqnarray}
%%%%%
the generalized Bell state $ | \phi_- ( N,0,r = 1 / \lambda ) \rangle $
and the squeezed vacuum state $ | \lambda \rangle $
are taken as the input state and EPR resource, respectively.
The matrix representing the input state is given by
%%%%%
\begin{eqnarray}
{\bf c}^{\rm in}
= {\bf E} [ | \phi_- ( N,0,r = 1 / \lambda ) \rangle ] .
\nonumber
\end{eqnarray}
%%%%%
Then, the fidelity is calculated for example with $ \lambda = 1/4 $ as
%%%%%
\begin{eqnarray}
&{}& \Delta F_{\rm MSV} \equiv 1 - F_{\rm MSV} [ N = 1 ; \lambda = 1/4 ] :
\nonumber \\
&{}& \left[ \begin{array}{ccc}
a & f^{(a,0)}{\delta \eta}^a
& f^{(a,1)} \nu {\delta \eta}^a \\
0 & - \! \! \! - \! \! \! - & 0 \\
1 & 1/8 & 0 \\
2 & - 1/256 & - 1/8 \\
3 & 0 & - 7/64 \\
4 & 0 & - 225/2048
\end{array} \right] \ ,
\nonumber
\end{eqnarray}
%%%%%
%%%%%
\begin{eqnarray}
&{}& \Delta F_{\rm MSV} \equiv 1 - F_{\rm MSV} [ N = 2 ; \lambda = 1/4 ] :
\nonumber \\
&{}& \left[ \begin{array}{ccc}
a & f^{(a,0)}{\delta \eta}^a
& f^{(a,1)} \nu {\delta \eta}^a \\
0 & - \! \! \! - \! \! \! - & 0 \\
1 & 35/192 & 0 \\
2 & - 391/36864 & - 1351/4608 \\
3 & - 77/786432 & - 107425/442368 \\
4 & 8473/226492416 & - 4611707/18874368
\end{array} \right] \ .
\nonumber
\end{eqnarray}
%%%%%

Numerical estimates are shown in Figs. \ref{msvn1} and \ref{msvn2}
for $ N = 1 $ and $ N = 2 $, respectively,
depending on $ \eta $ with $ \nu = 0 $ for simplicity,
where some typical values are taken for the relevant squeezing parameter
$ \lambda $.
The contributions of the dark count are actually negligible
for $ \nu \sim 10^{-4} $,
since $ f^{(0,1)} = f^{(1,1)} = 0 $ incidentally,
as seen in the above lists.
A fairly high fidelity $ F_{\rm MSV} > 0.94 $ can really be achieved
for $ \eta \geq 0.7 $ with small enough $ \lambda \leq 1/4 $.
The input generalized Bell state
$ | \phi_- ( N,0,r = 1 / \lambda ) \rangle $
may be prepared from a pair of squeezed vacuum states
$ | \lambda^\prime \rangle $ and $ | \lambda^{\prime \prime} \rangle $
with $ r = \lambda^{\prime \prime} / \lambda^\prime = 1 / \lambda $.
As seen so far, a high fidelity $ F_{\rm GB} > 0.95 $ can be achieved
for $ \eta \geq 0.7 $ with $ \lambda^{\prime \prime} \leq 1/4 $.
Then, the actual net fidelity to prepare
the truncated maximally squeezed vacuum state
$ | \lambda = 1 , N = 1,2 \rangle $ can be as high as 0.9, e.g.,
for $ \eta = 0.7 $ with $ \lambda = 1/4 $,
$ \lambda^\prime = (1/4)^2 $ and $ \lambda^{\prime \prime} = 1/4 $.

It should be remarked here that the input state and EPR state
may be exchanged in the preparation of $ | \lambda = 1 , N \rangle $.
Then, the fidelity somewhat changes
since the optical setup of Bell state detector
is asymmetric under the exchange of the input modes 1 and 2,
as explained before.
In fact, we have $ F_{\rm MSV} = 1 - f^{(0,1)} \nu + \ldots $
($ f^{(0,1)} \sim 10 - 50 $)
with $ | \psi_{\rm in}^{(2)} \rangle = | \lambda \rangle $
and $ | {\rm EPR} \rangle = | \phi_- ( N,0,r = 1 / \lambda ) \rangle $
for both the cases of $ N = 1 , 2 $.
It is really checked numerically
that the corrections of the order of $ \nu^0 $ are zero
up to $ {\delta \eta}^4 $ independent of $ \lambda $.
This case may be more favorable
since the fidelity is rather insensitive to $ \eta $.
Furthermore, a somewhat large $ \lambda $ may be taken
to increase the success probability.
The effect of the dark count is small enough for $ \nu \sim 10^{-4} $
with $ f^{(0,1)} \sim 50 $.
In any case, the fidelity for the preparation
of $ | \phi_- ( N,0,r = 1 / \lambda ) \rangle $ should be considered.

The net success probability to prepare $  | \lambda = 1 , N \rangle  $
from $ | \phi_- ( N,0,r = 1 / \lambda ) \rangle $
and $ | \lambda \rangle $ is estimated roughly \cite{KY-2003} as
%%%%%
\begin{eqnarray}
P_{\rm MSV} (N) \sim \frac{p(N)^2}{( N + 1 )^2} \lambda^{\prime 2N} ,
\nonumber
\end{eqnarray}
%%%%%
where $ \lambda^\prime = \lambda \lambda^{\prime \prime} $
with $ \lambda^2 , \lambda^{\prime 2} , \lambda^{\prime \prime 2} \ll 1 $.
Numerically, for example $ P_{\rm MSV} (1) \sim 4 \times 10^{-3} $
and $ P_{\rm MSV} (2) \sim 7 \times 10^{-6} $
with $ \lambda = 1/2 $, $ \lambda^\prime = 1/8 $
and $ \lambda^{\prime \prime} = 1/4 $.

%%%%%%%%%%
\section{Summary}
\label{sec:summary}
%%%%%%%%%%

In summary, we have analyzed the linear optical realization
of number-sum Bell measurement and number-state manipulation
by taking into account the realistic experimental situation,
specifically imperfectness of single-photon detector.
The present scheme for number-state manipulation
is based on the number-sum Bell measurement,
which is implemented with linear optical elements, i.e.,
beam splitters, phase shifters and zero-one-photon detectors.
Squeezed vacuum states and coherent states are used as optical sources,
while single-photon sources may not be required.
The linear optical Bell state detector has been formulated
quantum theoretically with a probability operator measure.
Then, the fidelity of manipulation and preparation
of number-states, particularly for qubits and qutrits,
has been evaluated in terms of the quantum efficiency $ \eta $
and dark count $ \nu $ of single-photon detector.
It will be encouraging for future experimental attempts
that a high fidelity is achievable
for $ \eta \gtrsim 0.7 $ and $ \nu \sim 10^{-4} $
with small enough squeezing parameters $ \sim 0.1 $
and coherent state amplitudes $ \lesssim 1 $.

\acknowledgments

The authors would like to thank K. Ogure, M. Senami and M. Sasaki 
for valuable discussions.

\end{document}